\documentclass[prb,twocolumn,amsmath,amssymb,aps]{revtex4}
\usepackage{graphicx}
\usepackage{bm}

\newcommand{\be}{\begin{equation}}
\newcommand{\ee}{\end{equation}}
\newcommand{\beq}{\begin{eqnarray}}
\newcommand{\eeq}{\end{eqnarray}}

\begin{document}

\title{Unzipping Vortices in Type-II Superconductors}

\author{Yariv Kafri$^1$, David R. Nelson$^2$ and Anatoli Polkovnikov$^{3}$}

\affiliation{$^1$ Department of Physics, Technion, Haifa 32000, Israel.\\
             $^2$ Lyman Laboratory of Physics, Harvard University, Cambridge, MA
  02138\\
  $^3$Department of Physics, Boston University, Boston, MA 02215 }

\date{\today}

\begin{abstract}
The unzipping of vortex lines using magnetic-force microscopy from
extended defects is studied theoretically. We study both the
unzipping isolated vortex from common defects, such as columnar pins
and twin-planes, and the unzipping of a vortex from a plane in the
presence of other vortices. We show, using analytic and numerical
methods, that the universal properties of the unzipping transition
of a single vortex depend only on the dimensionality of the defect
in the presence and absence of disorder. For the unzipping of a
vortex from a plane populated with many vortices is shown to be very
sensitive to the properties of the vortices in the two-dimensional
plane. In particular such unzipping experiments can be used to
measure the ``Luttinger liquid parameter'' of the vortices in the
plane. In addition we suggest a method for measuring the line
tension of the vortex directly using the experiments.
\end{abstract}

\maketitle

\section{Introduction}

The competition between thermal fluctuations, pinning and
interactions between vortices  leads to many novel physical
phenomena in type-II high-temperature
superconductors~\cite{blatter}. Examples include the melting of the
Abrikosov flux-lattice into an entangled vortex-liquid~\cite{ns} and
the proposed existence of low temperature Bose-glass~\cite{nv},
vortex glass~\cite{fisher_glass} and Bragg glass~\cite{nat_sch}
phases.

Many experimental probes have been used to study these phenomena.
They include decoration, transport and magnetization measurements,
neutron scattering, electron microscopy, electron holography and
Hall probe microscopes. More recently it has become possible to
manipulate single vortices, for example using magnetic force
microscopy (MFM)~\cite{Wadas92}. These can, in principle, measure
directly many microscopic properties which have been up to now under
debate or assumed. The possibility of performing such experiments is
similar in spirit to single molecule experiments on motor proteins,
DNA, and RNA which have opened a window on phenomena inaccessible
via traditional bulk biochemistry experiments~\cite{singlemol}.

In this spirit Olson-Reichhardt and Hastings~\cite{ORH04} have
proposed using MFM to wind two vortices around each other. Such an
experiment allows direct probing of the energetic barrier for two
vortices to cut through each other. A high barrier for flux lines
crossing has important consequences for the dynamics of the
entangled vortex phase.

In this paper we introduce and study several experiments in which a
single vortex is depinned from extended defects using, for example,
MFM. A brief account of the results can be found in
Ref.~[\onlinecite{knp}]. First we consider a setup where MFM is used
to pull an isolated vortex bound to common extended defects such as
a columnar pin, screw dislocation, or a twin plane in the presence
of point disorder. Using a scaling argument, supported by numerical
and rigorous analytical results, we derive the displacement of the
vortex as a function of the force exerted by the tip of a magnetic
force microscope. We focus on the behavior near the depinning
transition and consider an arbitrary dimension $d$. We argue that
the transition can be characterized by a universal critical
exponent, which depends {\it only on the dimensionality of the
defect}. We show that unzipping experiments from a twin plane
directly measures the free-energy fluctuations of a vortex in the
presence of point disorder in $d=1+1$ dimensions. To the best of our
knowledge, there is only one, indirect, measurement of this
important quantity in Ref.~[\onlinecite{Bolle}]. The form of the
phase diagram in the force temperature plane is also analyzed in
different dimensions. Related results apply when a tilted magnetic
field is used to tear away vortex lines in the presence of point
disorder, which was not considered in earlier work on clean
systems.~\cite{hatano}. Furthermore, we show that a possible
experimental application of the scaling argument is a direct measurement of the vortex line tension in an unzipping
experiment. As we will show in this paper, in a system of finite size, the displacement of the flux line at the transition
depends only on the critical force exerted on the flux line by the
MFM tip, the flux line tension and the sample thickness. Thus
unzipping experiments can provide valuable information on the
microscopic properties of flux lines.

Next we consider a setup where a single vortex is pulled out of a
plane with many vortices. It is known that the large-scale behavior
of vortices in a plane is characterized by a single dimensionless
number, often referred to as the Luttinger liquid parameter due to
an analogy with bosons in $d=1+1$ dimensions. We show that
experiments which unzip a single vortex out of the plane can be used
to directly probe the Luttinger liquid parameter. We also discuss
the effects of disorder both within the defect and in the bulk with the same setup.

\section{Unzipping a vortex from a defect}
\label{Sec2}

\subsection{Review of clean case}
\label{sectioncleancase}

\begin{figure}[ht]
\center
\includegraphics[width=8cm]{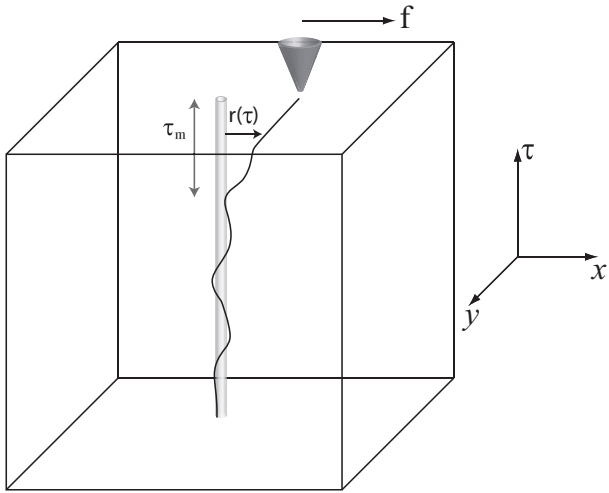}
\caption{A MFM tip applies a constant force ${\bf f}$ which pulls the
vortex away from the defect. The configuration of the vortex is
represented by ${\bf r}(\tau)$. We assume throughout that the vortex
is always bound to the defect at the bottom of the sample so that
${\bf r}(\tau=0)=0$. }\label{fig:clean_unzip}
\end{figure}

We begin by considering the unzipping of a vortex from an extended
defect in a clean sample. For a columnar defect the system is
depicted in Fig.~\ref{fig:clean_unzip}. At the top of the sample the
MFM applies a constant force ${\bf f}$ which pulls the vortex away
from the defect. We assume that at the bottom of the sample the
vortex is always bound to the defect at a specific location. This
assumption will not influence the results since below the unzipping
transition the flux line is unaffected by the boundary conditions at
the far end of the sample.

In the absence of external force, and for an external field aligned
with the defect, the appropriate energy for a given configuration
${\bf r}(\tau)$ of the vortex is given by \cite{blatter}:
\begin{equation}
F_0\!\!=\!\!\!\int_0^L \!\!\!\!d \tau \left[ \frac{\gamma}{2}
(\partial_\tau {\bf r}(\tau))^2+V({\bf r}(\tau)) \right] .
\label{f0}
\end{equation}
Here $\gamma$ is the line tension and $L$ is the length of the
sample along the $\tau$ direction. The vector ${\bf r}(\tau)$
represents the configuration of the vortex  in the $d$ dimensional
space and $V({\bf r})$ is a short-ranged attractive
potential describing the $d'$-dimensional extended defect (in
Fig.~\ref{fig:clean_unzip} $d=3$ and $d'=1$). The effect of the the
external force, exerted by the MFM, can be incorporated by adding to
the free energy the contribution
\begin{equation}
F_1=-{\bf f}\cdot {\bf r(L)}=-\int_0^L {\bf f}\cdot
\partial_\tau \bf r(\tau)\,d\tau
\label{eq:unzipfe}
\end{equation}
where we have used ${\bf r}(\tau=0)={\bf 0}$. Here ${\bf f}$
stands for the local force exerted by the MFM in the transverse
direction. The free energy of a given configuration of the vortex
is given by
\begin{equation}
F({\bf r})=F_0({\bf r})+F_1({\bf r}) \;.
\end{equation}
The problem, as stated, has been studied first in the context of
vortices in the presence of a tilted magnetic field~\cite{hatano}
and the results have been applied to the related problem of DNA
unzipping~\cite{Lubensky}.

\begin{figure}[ht]
\center
\includegraphics[width=8cm]{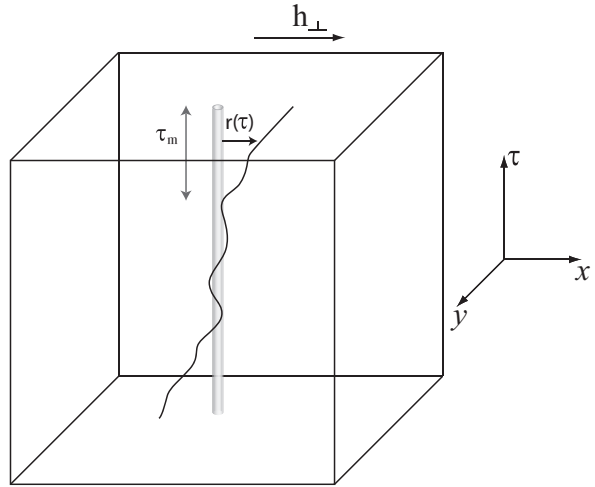}
\caption{Same as in Fig.~\ref{fig:clean_unzip} but with a transverse
magnetic field instead of the MFM force tearing the flux line away from a
defect.}\label{fig:clean_unzip_mag}
\end{figure}

We note that a similar setup can be
achieved by using a transverse magnetic field instead of the
external force. See Fig.~(\ref{fig:clean_unzip_mag}).
Indeed in the free energy (\ref{eq:unzipfe}) the external force
couples to the slope of the flux line $\partial_\tau {\bf r}$ in the
same way as the external magnetic field does~\cite{hatano}. The only
difference between the two setups is that there are now equal and
opposite forces acting on the top and bottom ends of the sample.
However this difference is important only in short samples, where
the two ends of the flux line are not independent from each other.

In this paper we focus on the thermal average of distance of the tip
of the vortex from the extended defect $\langle x_m (\tau=L)\rangle$. This quantity
is related to the thermal average of the length of the vortex that is unzipped from the
defect, $\langle \tau_m \rangle$, through $\langle x_m \rangle = f
\langle \tau_m \rangle / \gamma$. Here and throughout the paper
$\langle \ldots \rangle$ denotes a thermal average while an overbar
denotes an average over realizations of the disorder.

As stated above the universal behavior of $\langle \tau_m \rangle$
(or equivalently $\langle x_m \rangle$) within this disorder-free model have been
derived previously. Here we sketch two approaches which will be
generalized to samples with quenched disorder in the rest of the paper.

In the first approach, instead of directly summing over
configurations of the vortex we perform the sum in two parts by
dividing the vortex into bound and unbound segments. The unbound
segment starts at the point where the vortex departs the defect
without ever returning to hit it again up to the top of the sample.
Using Eq.~(\ref{eq:unzipfe}) it is straightforward to integrate over
vortex configurations to obtain for the partition function of the
unzipped segment
\begin{eqnarray}
Z_u(\tau_m)&=&\int {\cal D}{\bf r}(\tau)\,\mathrm
e^{-\beta\int_0^{\tau_m}\! d \tau \left[ \frac{\gamma}{2}
(\partial_\tau {\bf r}(\tau))^2-{\bf
f}\cdot\partial_\tau{\bf r(\tau)}\right]} \nonumber \\
&\propto& \mathrm e^{\tau_m \beta f^2 /2\gamma} \;,
\label{free_unzip1}
\end{eqnarray}
so that the free energy associated with this conditional partition function is
\begin{equation}
{\cal F}_u(\tau_m)=-\beta^{-1}\ln Z_u(\tau_m)= -  f^2 \tau_m/
2\gamma \;,
\label{free_unzip}
\end{equation}
where $\beta$ is the inverse temperature. Henceforth in this paper we
set $\beta=1$, which can be always achieved by appropriate rescaling
of the energy units. Even though the above sum also runs over
configurations which return to the defect it is easy to verify that
these configurations give rise to exponentially small correction in
the $\tau_m$. Equation~(\ref{free_unzip}) implies that as the force,
${\bf f}$, increases the free energy density of the unzipped portion
of the vortex decreases. In contrast, the free energy density of the
bound part is, clearly, independent of the force and given by ${\cal
F}_b(\tau_m)=V_0(L-\tau_m)$, where $V_0$ is the free energy per unit
length of a bound vortex and $L$ is the length of the sample along
the defect. The vortex will be unzipped when $f=f_c=\sqrt{2 \gamma
|V_0|}$ such that the free-energy densities of the bound and
unzipped states are equal.

In this representation the total free energy of the vortex is
given by
\begin{equation}
{\cal F}(\tau_m)={\cal F}_u(\tau_m)+{\cal F}_b(\tau_m)\;.
\end{equation}
The unconstrained partition function of the model is given by
\begin{equation}
Z=\int_0^L d\tau_m e^{-(f_c^2-f^2)\tau_m/2 \gamma} \;.
\end{equation}
Since both results are independent of the dimensionality of the
defect (columnar or planar) near the transition one always finds
in the $L \to \infty$ limit
\begin{equation}
\langle \tau_m \rangle \sim \frac{1}{(f_c-f)^\nu} \;,
\label{eq:clean}
\end{equation}
with $\nu=1$. Note, that it can easily be seen that approaching the
transition from above the average length of the vortex which is
bound to the defect, $\langle (L-\tau_m) \rangle$, diverges in the
same manner~\cite{hatano}.

An alternative approach, which will also be useful in this paper,
uses the mapping of the problem to the physics of a fictitious
quantum particle \cite{NelsonBook}. The contribution of the external
field ${\bf f}$ to the free energy now manifests itself as an
imaginary vector potential acting on the particle in $d-1$
dimensions (with the $\tau$ axis acting as a time direction).
Explicitly, using the standard conversion from path-integrals (see
Ref.~[\onlinecite{hatano}] for details) one finds that the problem
can be described in terms of a non-Hermitian Hamiltonian:
\begin{equation}
{\cal H}={1\over 2\gamma} {\bf p}^2 -{i\over \gamma} {\bf f} \cdot
{\bf p} +V({\bf r}) \;,
\label{H}
\end{equation}
where ${\bf p}=\frac{1}{i} \vec{\nabla}$ is the momentum operator.
In this language the vortex is bound to the defect as long as there
is a bound state in the Hamiltonian. As mentioned above $i {\bf f}$
is equivalent to a constant imaginary vector potential. This analogy
makes it apparent that solutions of the non-Hermitian problem can be
related to those of the Hermitian Hamiltonian (where one sets ${\bf
f}=0$) by an imaginary gauge-transformation~\cite{hatano}. In
particular the left $\psi^L_n({\bf r},{\bf f})$ and the right
$\psi^R_n({\bf r},{\bf f})$ eigenfunctions of the non-Hermitian
problem can be obtained from those of the Hermitian problem,
$\psi_n({\bf r},{\bf f}={\bf 0})$, using
\begin{eqnarray}
\psi^R_n({\bf r},{\bf f}) &=&  {\cal U} \psi_n({\bf r},{\bf
f}={\bf 0}) \nonumber \\
\psi^L_n({\bf r},{\bf f}) &=&  \psi_n({\bf r},{\bf f}={\bf 0})
{\cal U}^{-1} \;, \label{eq:gauge1}
\end{eqnarray}
where
\begin{equation}
{\cal U}=\mathrm e^{{\bf f} \cdot {\bf r}}= \mathrm e^{f x}\; ;
\;\;\;\;\; {\cal U}^{-1}=\mathrm e^{-{\bf f} \cdot {\bf r}}=\mathrm
e^{-f x} \;.
\label{eq:gauge2}
\end{equation}
The universal behavior of $\tau_m$ at the transition,
Eq.~(\ref{eq:clean}), was obtained in Ref.~[\onlinecite{hatano}] by noting
that
\be
\langle\tau_m\rangle\propto \langle x_m\rangle ={\int x\psi^R_n({\bf
r}) d{\bf r}\over \int \psi^R_n({\bf r}) d{\bf r}}\propto{1\over
{f_c-f}}, \label{tau_m}
\ee
where $\psi^R_n({\bf r})\propto \mathrm e^{-f_c r}$ at long $r$. We
note that the imaginary gauge transformation is justified only at
$f<f_c$. Otherwise the integrals in Eq.~(\ref{tau_m}) formally
diverge. In principle this divergence can be fixed by proper
treatment of the boundary conditions~\cite{AHNS}.

Finally we discuss the phase diagram of the model. It is natural
that as the temperature rises the value of the critical force
needed for the unzipping decreases (at very low temperatures a
possible reentrance has been discussed in
literature~\cite{Reentrant,Reentrant2}). Here we focus on the
existence of a critical force for any strength of the attractive
potential. The question is completely equivalent to that of the
existence of a localized vortex line in the absence of the external force.
Using the analogy to the quantum mechanical problem it is well known
that in two dimensions (corresponding to a three dimensional sample)
and below, as long as $\int_{-\infty}^\infty d{\bf r} V({\bf r}) <0$
there exists a bound state. Therefore, in real samples, as long as
the potential satisfies this condition there is always a minimum nonzero
critical force required to unbind the flux line from the defect.

\subsection{Scaling arguments for the disordered case.}
\label{sectionscaling}

\subsubsection{Critical Behavior}
We now consider the effect of point
disorder. In this case the free energy of a given vortex
configuration without the contribution from the force exerted by
the MFM is given by
\begin{equation}
F_0\!\!=\!\!\!\int_0^L \!\!\!\!d \tau \left[ \frac{\gamma}{2}
(\partial_\tau {\bf r}(\tau))^2+V({\bf r}(\tau))+\mu({\bf
r}(\tau),\tau) \right] . \label{f0dis}
\end{equation}
The frozen point disorder $\mu({\bf r},\tau)$ is assumed to be uncorrelated
and Gaussian distributed with $\overline{\mu}=0$ and
$\overline{\mu({\bf r},\tau)\mu({\bf r'},\tau')}=\sigma \delta({\bf
r}-{\bf r'})\delta(\tau-\tau')$. As mentioned above the overbar denotes an average over realizations of the
disorder and the $\langle \dots\rangle$ denotes averaging over
thermal fluctuations. The contribution from the force exerted by the
MFM retains the same form as in Eq.~(\ref{eq:unzipfe}).

The free energy, Eq.~(\ref{f0dis}), without the contribution from
the external force has been studied previously with and without the
presence of an extended defect. Without the defect it is well known
that if one fixes one end of the vortex at the bottom of the sample
the mean-square displacement, $\langle x^2(\tau) \rangle$ of the
vortex after traveling a distance $\tau$ into the sample behaves as
$\langle x^2(\tau) \rangle = B\tau^{2\zeta(d)}$, where
$\zeta(d=2)=2/3$ and $\zeta(d=3) \approx 0.61$ is the wandering
exponent~\cite{Karreview}. Moreover, for a given realization of disorder there is
typically a single dominant trajectory of the vortex and
realizations with two competing minima are very rare. Finally, it is
known that the free energy fluctuations scale as
\begin{equation}
\delta {\cal F} \propto \tau^{\omega(d)} \label{eq:fefluct}
\end{equation}
with $\omega(d=2)=1/3$, $\omega(d=3) \approx 0.22$. The exponents
$\omega$ and $\zeta$ are not independent. They satisfy the
scaling relation: $\omega=2 \zeta -1$. Note that as long as there is
no defect the free energy fluctuations are unaltered even in the
presence of a force acting on the vortex because this force can be
simply gauged away.

The universal properties of the unzipping transition with point
disorder can be analyzed using a simple scaling argument adapted
from~[\onlinecite{Lubensky}] for the unzipping of DNA (for an
abbreviated account, see Ref.~[\onlinecite{knp}]).

Following the discussion in the previous section, which analyzed
the unzipping problem with no disorder, we sum over configurations
of the unzipped part of the vortex and the zipped part of the
vortex separately. As before we denote the free energy of the
unzipped segment by ${\cal F}_u$ and the free energy of the bound
segment by ${\cal F}_b$. The total free energy is then given by
\begin{equation}
{\cal F}(\tau_m)={\cal F}_u (\tau_m)+{\cal F}_b(L-\tau_m).
\label{fm}
\end{equation}
For large $L$ and $\tau_m$ we can rewrite
\be
\mathcal F_b(L-\tau_m)=\mathcal F_b(L)-\mathcal F_b(\tau_m).
\label{fb}
\ee
For a one dimensional defect, like a columnar pin, $\mathcal F_b(L)$
is a constant equal to the free energy of a flux line completely
localized on the pin. For higher dimensional defects like a twin
plane, there is a subtlety. To see this, consider a zero temperature
case first. Then the partition function is determined by the optimal
trajectory of a flux line (in the presence of point disorder)
starting at $\tau=0$ and ending at $\tau=L-\tau_m$. However, this
trajectory might not be a part of the optimal trajectory going all
the way from $\tau=0$ to $\tau=L$. In this case clearly the relation
(\ref{fb}) does not hold. Since this relation does not hold at zero
temperature, it should not hold at finite temperatures as well.
However, in Ref.~[\onlinecite{fisher}] it was argued that such
situations (where the optimal trajectories do not coincide) are very
rare and can be neglected \cite{footnote}. Thus ignoring the
constant term $\mathcal F_b(L)$ we can rewrite Eq.~(\ref{fm}) as
\begin{equation}
{\cal F}(\tau_m)={\cal F}_u(\tau_m)-\mathcal F_b(\tau_m).
\end{equation}
We can identify three contributions to the free energy above. The
first is due to the average free energy difference between a vortex
on the defect and in the bulk of the sample. Close to the
transition, similarly to the analysis in the clean case, it is
linear in $\tau_m$ and behaves as $a(f_c-f)\tau_m$, where $a\approx
f_c/\gamma$ is a positive constant. The second contribution $\delta
{\cal F}_{b}$, comes from the free-energy fluctuations arising from
that part of the point disorder which is localized on or near the
defect. For a columnar defect of dimensionality $d'=1$ this
contribution is due to the sum of the fluctuations in the free
energy about the mean. The central limit theorem implies that it
behaves as $\tau_m^{1/2}$ at large $\tau_m$. For $d'>1$ this result
is modified and one can use known results for the free energy of a
direct path in a random media (see Eq.~(\ref{eq:fefluct})), which
leads to $\delta {\cal F}_{b}\propto \tau_m^{\omega(d')}$, where
$d'$ is the dimensionality of the defect \cite{Karreview}. Finally,
there is a contribution to the free energy fluctuations from the
interaction of the unzipped part of the vortex with the bulk point
disorder, $\delta {\cal F}_{u}$. This contribution behaves similarly
to $\delta \mathcal F_b$ with a different bulk exponent: $\delta
{\cal F}_{u}\propto \tau_m^{\omega(d)}$, where $d>d'$ is the
dimensionality of the sample. Collecting all three terms gives:
\begin{equation}
{\cal F}(\tau_m)=a(f_c-f)\tau_m +\delta\mathcal F_{b}(\tau_m)
+\delta\mathcal F_u(\tau_m)\;.
\label{f}
\end{equation}
As discussed above, $\omega(d)$ has been studied extensively in the
past and it is well known that $\omega(d')>\omega(d)$ for any
$d'<d$~\cite{Karreview}. Therefore, disorder on or close to the
defect controls the unbinding transition for {\it any dimension}
when $\tau_m$ is large and the problem is equivalent to unzipping
from a sample with point disorder localized on the defect. In
particular, this result implies that unzipping from a columnar
defect with point disorder in the bulk is in the same universality
class as unzipping of a DNA molecule with a random sequence of base
pairing energies. In fact, point disorder is likely to be
concentrated within real twin planes and near columnar damage tracks
created by heavy ion irradiation and near screw dislocations,
strengthening even more the conclusions of this simple argument.

The disorder averaged partition function is dominated by the minimum
of the free energy and thus by configurations with $\delta \mathcal
F_b(\tau_m)<0$. Using ${\cal F}(\tau_m)\approx a(f_c-f)\tau_m -b
m^{\omega(d')}$, where $b$ is a positive constant, the partition
function
\begin{equation}
Z=\int_0^L d\tau_m e^{-{\cal F}(\tau_m)}
\end{equation}
can be evaluated using a saddle-point approximation. We then find the
value of $\tau_m$ at the saddle point satisfies $a(f_c-f) =
\omega(d') b \tau_m^{\omega(d')-1}$. Therefore
\begin{equation}
\overline{\langle \tau_m \rangle} \sim \frac{1}{(f_c-f)^{\nu}}, \;\;
\nu=[1-\omega(d')]^{-1} \;,
\label{nu}
\end{equation}
with
\begin{eqnarray}
\nu&=&2 \;\;\;\;\;\;\;\, {\rm for} \;\;\; d'=1 \nonumber \\
\nu&=&3/2 \;\;\;\; {\rm for} \;\;\; d'=2\;.
\label{valuenu}
\end{eqnarray}
The result for the columnar defect ($d'=1$) agrees with known results from
DNA unzipping~\cite{Lubensky}.

To check the scaling argument we have preformed numerical
simulations in $d=1+1$ dimensions and have been able to solve analytically the
closely related problem of vortex unzipping from a wall in $d=1+1$ dimensions. We
have also studied the simplified problems of unzipping from a $d'=1$
and $d'=2$ dimensional defect with disorder localized on the defect.
While the $d'=1$ problem was solved previously~\cite{oper,Lubensky}
below (in Sec.~\ref{replica}) we describe a new method applying the replica method. Using
some approximations,this approach can be generalized to a $d'=2$
dimensional defect. In the next sections we describe all these
results which support the simple scaling argument presented above.

\subsubsection{Finite size scaling and determination of the flux
line tension.}

Another consequence of the result (\ref{nu}) is the possibility to
perform a finite size scaling analysis~\cite{privman}. In
particular, near the unzipping transition the unzipping length
$\overline{\langle \tau_m\rangle}$ should be a function of the
reduced force $\epsilon \propto (f_c-f)$ and the system size $L$. If
the scaling ansatz (\ref{nu}) is correct then we must have
\be
\overline{\langle \tau_m\rangle}/L= g(\epsilon L^{1/\nu}),
\label{scaling1}
\ee
where $g(x)$ is some scaling function. Quite generally we expect
that when $x\gg 1$ finite size effects are unimportant and $g(x)
\sim 1/x^\nu$ so that we recover the scaling (\ref{nu}) and at $x\ll
1$ we have $g(x)\to g_0$, where $g_0$ is a constant. Note that the
constant $g_0$ does not depend on the system size. This constant is
a universal number of the order of one, which depends on the
dimensionality of the defect. As we find below for the unzipping
from a columnar pin $g_0=0.5$ and for the unzipping from a twin
plane $g_0\approx 0.7$. Relation (\ref{scaling1}) can be used to
extract the line tension $\gamma$ through:
\be
\gamma=f_c g_0 {L\over \langle x_m\rangle}.
\label{gamma}
\ee
Here $f_c$ is the critical force and $\langle x_m\rangle$ is the
displacement of the MFM tip at the transition. To derive
Eq.~(\ref{gamma}) we used the fact that for the unbound
segment $\langle x_m\rangle=\overline{\langle\tau_m\rangle}
f/\gamma$.

We emphasize that the unzipping transition is first order both in
the clean and disordered cases. Indeed, the unzipping occurs only at
the boundary and does not affect total free energy of the flux line
in the thermodynamic limit. Nevertheless, similar to wetting
phenomena near first order transitions, this transition possesses
scaling properties characteristic of second order transitions like
diverging correlation lengths, finite size scaling etc.

\subsubsection{Phase diagram}
\label{scalingphasediagram}

Next, we use scaling arguments to consider the behavior of the
critical force as the strength of the disorder is varied. In
particular we focus on the existence of a critical strength of the
disorder beyond which the flux line spontaneously unzips without any
external force. The disorder induced unbinding transition of the
vortex from the defect in the absence of the force has been
considered previously~\cite{Kar85,Zap91,Tang93,Bal93,Kol92,Bal94}.
Below we extend a scaling argument presented first by Hwa and
Nattermann in Ref.~[\onlinecite{HwaNatter}] for a columnar pin in arbitrary
dimensions to include also planar defects.

Assume that the vortex is localized within a distance $l_\perp$ from
a columnar pin or a twin plane. Then, it consists of uncorrelated
segments of length $l_\parallel$ related to $l_\perp$ via
\begin{equation}
l_\parallel \propto  l_\perp^{1/\zeta} \;,
\end{equation}
where $\zeta$ is the wandering exponent defined above. Each of these
segments has a free energy excess of order $l_\parallel^\omega$
higher than the energy of the delocalized vortex. The free energy cost per
length of localization therefore scales as $l_\parallel^{\omega-1}
\sim l_\perp^{(\omega-1)/\zeta}$. Clearly, a strong enough pinning
potential gives rise to a constant energy gain per unit length,
which suppresses the random energy cost of localization (note that
$\omega<1$ in any dimension).

So far we established that a localized phase can exist. Now let us
consider perturbative effects of a weak attractive potential and ask
whether the vortex immediately becomes bound to the defect. The free
energy gained in the presence of the defect, $\delta F$, can be
inferred by perturbations in the strength of the defect pinning energy $V$.
Assume that one end of the vortex is held on the defect. Then the
energy gain due to the attractive potential by the defect is
associated with the the return probability of the flux line back to
the defect. Since the root mean square displacement behaves as
$l_\parallel^\zeta$  the return probability behaves as
$l_\parallel^{1-(d-d^\prime)\zeta}$. The free energy gained by
hitting the defect $\delta F$  therefore scales as
$l_\parallel^{1-(d-d^\prime)\zeta}$.

To determine if the pin is relevant one has to compare this energy
to the intrinsic variations in the free energy, $\Delta F$, which
scale as $l_\parallel^{\omega(d)}$. This yields
\begin{equation}
g= \frac{\delta F}{\Delta F} \propto l_\parallel^{\varepsilon}
\end{equation}
with
\begin{equation}
\varepsilon=1-\zeta(d-d^\prime)-\omega=2-(d+2-d^\prime)\zeta.
\end{equation}
When $\varepsilon<0$ the defect potential is irrelevant, i. e. the
system gains more energy by minimizing returns to the defect, while
if $\varepsilon>0$ it is relevant, i. e. long excursions are
energetically costly.

As mentioned above in $d=3$ numerical simulations indicate that
$\zeta \approx 0.6$ which gives for the planar defect ($d^\prime=2$)
$\varepsilon \approx 1/8$ and for the columnar pin ($d^\prime=1$)
$\varepsilon \approx -1/2$. Therefore, a weak twin plane is always relevant and
the vortex is always bound to it. However, a weak columnar pin is
irrelevant and then one expects an unbinding transition. In $d=2$,
where there can be no twin plane, $\zeta=2/3$ and the columnar
defect is found to be marginal. As argued in Ref.~[\onlinecite{HwaNatter}]
it is in fact marginally {\it relevant}.

To summarize this discussion for columnar defects in 3 dimensional samples we
expect there is a critical strength of the bulk disorder beyond which the
flux line spontaneously unzips even at zero force. In contrast for a planar defect in
3 dimensions and for columnar defects in planar 2 dimensions superconductors we expect
that for any strength of the disorder there is a finite non-zero
value of the force needed to unzip the vortex.

Next, we will check the scaling (\ref{nu}) and the anticipated
localization / delocalization behavior for a number of different
situations using both analytical methods based on the replica trick
and numerical simulations.

\subsection{Unzipping from a disordered columnar pin without excursions.}
\label{replica}

We start our quantitative analysis from the simplest situation,
where one can get exact analytical results. Namely, we consider
unzipping from a 1D pin with disorder localized only on the pin.
Additionally we neglect all excursions of the vortex line from the
pin except for the unzipped region. This problem then becomes
identical to DNA unzipping. In Ref.~[\onlinecite{Lubensky}] the
authors analyzed this problem using a Fokker-Planck approach and
indeed derived $\nu=2$ near the unzipping transition. Here we show
how the same problem can be solved using the replica trick. The
solution was sketched  in Ref.~[\onlinecite{kp}]. Here we review the
derivation for completeness and provide additional details.

Ignoring excursions of the bound part of the flux line into the bulk
gives the free energy a particularly simple form. We again write it
as a sum over the contribution from the bound and unbound segments.
The bound segment contribution is given by ${\cal
F}_b(\tau_m)=V_0(L-\tau_m)+\int_{\tau_m}^L d \tau_m' U(\tau_m')$,
where $V_0<0$ is the mean value of the attractive potential, $L$ is
the length of the columnar defect which is assumed to be very large,
and $U(\tau_m)$ is a random Gaussian uncorrelated potential with
zero mean satisfying
$\overline{U(\tau_{m_1})U(\tau_{m_2})}=\Delta\delta(\tau_{m_1}-\tau_{m_2})$.
The contribution from the unzipped part takes the same form as in
the clean case (see Eq. (\ref{free_unzip})). Collecting the two
terms gives:
\begin{equation}
\mathcal F(\tau_m)=\epsilon \tau_m+\int_{\tau_m}^L d \tau_m'
U(\tau_m').
\label{fz}
\end{equation}
As before we work in the units, where $k_B T=1$. In the equation
above the deviation from the unzipping transition is measured by
$\epsilon=(f_c^2-f^2)/2\gamma$, where $f$ is the force applied to
the end of the flux line and $f_c=\sqrt{2\gamma |V_0|}$ is the
critical force. In Eq.~(\ref{fz}) we dropped an unimportant constant
additive term $V_0 L$.

The statistical properties of the unzipping transition can be
obtained by considering $n$ replicas of the partition function $Z(\tau)=\exp(-\mathcal
F(\tau))$~\cite{edwards anderson}:
\begin{equation}
\overline{Z^n}=\int_0^L d\tau_1\ldots\int_0^L
d\tau_n\,\overline{\exp\left(-\sum_{\alpha=1}^n \mathcal
F(\tau_\alpha)\right)},
\label{Z_n}
\end{equation}
where the overbar denotes averaging over point disorder. The
averaging procedure can be easily done for a positive integer $n$.
We eventually wish to take the limit $n \to 0$. First we order the
coordinates $\tau_j$, where the $j^{th}$ replica unbinds from the
pin according to: $0\leq \tau_1\leq \tau_{2}\leq\dots\leq \tau_n$.
Then for $\tau\in[0,z_1)$ there are no replicas bound to the
columnar pin, for $\tau \in[\tau_1,\tau_2)$ there is one replica on
the pin until finally for $L \geq \tau\geq \tau_n$ all $n$ replicas
are bound to the pin. Using this observation and explicitly
averaging over the point disorder in Eq.~(\ref{Z_n}) we arrive at:
\be
\overline{Z^n}\!=n!\!\int\limits_0^L d\tau_1.\,.\!\!\int\limits_{\tau_{n-1}}^L
\!\!d\tau_n\exp\!\left[-\!\sum\limits_{j=1}^n\! \epsilon \tau_j+
{\Delta\over 2} j^2(\tau_{j+1}-\tau_{j})\right],
\label{tuam2}
\ee
where we use the convention $\tau_{n+1}=L$. The integral above is
straightforward to evaluate in the $L \to \infty$ limit so that
\begin{eqnarray}
&&\overline{Z^n}=\mathrm e^{n^2L\Delta/2}{1\over \epsilon_n^n}\prod_{j=1}^n
{1\over 1-\kappa_n j} \nonumber
\\
&&=\mathrm
e^{n^2L\Delta/2}\left({2\over\Delta}\right)^n
{\Gamma(1+1/\kappa_n-n)\over\Gamma(1+1/\kappa_n)}
\;, \phantom{XXX} \label{eq:partunzip}
\end{eqnarray}
where $\epsilon_n=\epsilon+\Delta n$ and
$\kappa_n=\Delta/2\epsilon_n$. The exponential prefactor is an
unimportant overall contribution of the whole columnar pin while the
rest of the expression is the ($L$ independent) contribution from
the unzipped region. Interestingly the restricted partition
functions for the unbinding problem from a hard wall (with no
external force) and for the unzipping from a 1 dimensional pin are identical
and thus there is equivalence between the two problems (see
Ref.~[\onlinecite{kp}] for more details.)

The disorder-averaged free energy is given by the limit
$\overline{\mathcal F}=-\lim_{n \to 0}
(\overline{Z^n}-1)/n$~[\onlinecite{edwards anderson}]. With the help
of Eq.~(\ref{eq:partunzip}) one obtains
\begin{equation}
\overline{\mathcal F}=\ln (\epsilon \kappa) + \Psi(1/\kappa),
\label{free_en}
\end{equation}
where $\Psi(x)$ is the digamma function and
$\kappa=\Delta/2\epsilon$. The unzipping transition occurs at
$\epsilon=0$ or equivalently at $\kappa \to \infty$. The expression
(\ref{free_en}) is identical to the one found in
Ref.~[\onlinecite{oper}] using a Fokker-Planck equation approach,
supporting the validity of the analytic continuation in $n$ for this
particular application of the replica calculation.

It is easy to see that this free energy yields
\begin{equation}
\overline{\langle \tau_m\rangle}={\partial \overline{\mathcal F}\over
\partial\epsilon}={1 \over \kappa\epsilon}\Psi^{(1)}(1/\kappa),
\label{zav}
\end{equation}
where $\Psi^{(n)}(x)$ stands for the $n$-th derivative of the
digamma function. The expression above predicts a crossover from
$\overline{\langle \tau_m\rangle}\approx 1/\epsilon$ for $\kappa\ll
1$ (far from the transition) to $\overline{\langle
\tau_m\rangle}\approx\kappa/\epsilon=\Delta/\epsilon^2$ for
$\kappa\gg 1$ (close to the transition) similarly to the unzipping
from the wall problem analyzed above. Also, it is easy to check that
\begin{equation}
w=\overline{\langle \tau_m^2 \rangle - \langle \tau_m
\rangle^2}={\partial^2 \overline{\mathcal F}\over \partial\epsilon^2}=-{1 \over
(\kappa\epsilon)^2}\Psi^{(2)}(1/\kappa). \label{fav}
\end{equation}
Here there is a crossover from $w \approx 1/\epsilon^2$ for $\kappa
\ll 1$ to $w \approx 2 \kappa/\epsilon^2=\Delta/\epsilon^3$ for
$\kappa\gg 1$. As has been noted in the context of DNA unzipping
\cite{Lubensky} $\sqrt{w}/\overline{\langle \tau_m\rangle}$ changes
from being of order unity for the weakly disordered $\kappa \ll 1$ case to
$\sim \epsilon^{1/2}$ for $\kappa \gg 1$. Thus for $\kappa \gg 1$,
close to the unzipping transition, thermal fluctuations become
negligible and one can work in the zero temperature limit.

The simplicity of the problem also allows finding the higher moments
of the distribution. Here we evaluate the second moment, which gives
the width of the distribution of $\overline{\langle \tau_m\rangle}$
due to different disorder realizations. Note that since the order of
averaging over thermal fluctuations and disorder is important this
quantity can not be extracted directly from Eq. (\ref{fav}). To
proceed we consider the generating function, ${\cal W}_n(\epsilon_j)$ defined by
\be
{\cal W}_n(\epsilon_j)=\int\limits_{0}^L
d\tau_1\ldots\int\limits_{\tau_{n-1}}^L d\tau_n\,\mathrm
e^{-\sum\limits_{j=1}^n \epsilon_j\tau_j+\Delta/2
j^2(\tau_{j+1}-\tau_j)}\!\!. \nonumber \label{zm1}
\ee
The second (and similarly the higher) moments can be found by
differentiating ${\cal W}_n$ with respect to $\epsilon_j$:
\begin{equation}
\overline{\langle \tau_m^2\rangle}=\lim_{n\to 0} \left. {1\over
{\cal W}_n(\epsilon_j)}\,{1\over n}\sum_{j=1}^n {\partial^2 {\cal
W}_n(\epsilon_j)\over\partial
\epsilon_j^2}\right|_{\epsilon_j=\epsilon}. \label{zm3}
\end{equation}
Upon evaluating the integral, we find
\begin{equation}
{\cal W}_n(\epsilon_j)=\prod_{j=1}^n {1\over
\sum_{k=1}^j\epsilon_k\,-\,\Delta j^2/2} \label{zm2}
\end{equation}
and correspondingly
\begin{equation}
\overline {\langle \tau_m^2\rangle}={1\over \epsilon^2}\lim_{n\to
0}{1\over n}\sum_{j=1}^n {2\over 1-\kappa j}\sum_{k=j}^n {1\over k
(1-\kappa k)}.
\end{equation}
This double sum can be calculated using a trick similar to the one
described in Ref.~[\onlinecite{Kardar}]:
\begin{eqnarray}
\overline{\langle \tau_m^2\rangle}&=&{2\kappa^2\over \epsilon
^2}\int\!\!\!\!\!\!\int\limits_{\!\!\!\!x>y>0}\!\!\!\! dx dy
{1\over \mathrm e^{\kappa x}-1}{y\,\mathrm e^{-y}\over \mathrm
e^{\kappa y }-1}\left[ \mathrm e^{\kappa y}+\mathrm e^{2y}\mathrm
e^{\kappa
x-x}\right]\nonumber\\
&-&{4\over \kappa
\epsilon^2}\Psi^{(1)}(1/\kappa)\left(C+\Psi(1/\kappa)\right),
\label{z2}
\end{eqnarray}
where $C\approx 0.577$ is Euler's constant. In the limit of weak
disorder or high temperature $\kappa\ll 1$, not surprisingly, we get
$\overline{\langle \tau_m^2\rangle }\approx 2/ \epsilon^2$, which
agrees with the Poissonian statistics of $\tau_m$ with an average
given by $\overline{\langle \tau_m \rangle}=1/\epsilon$. In the
opposite limit $\kappa\gg 1$ one finds $\overline{\langle
\tau_m^2\rangle }=4\kappa^2/ \epsilon^2$. Note that
$\overline{\langle \tau_m\rangle}=\kappa/\epsilon$, thus the
relative width of the distribution ($\delta \tau_m/\overline{\langle
\tau_m\rangle}$), defined as the ratio of the variance of the
unzipping length $\tau_m$ to its mean is larger by a factor of
$\sqrt{3}$ than that in the high temperature regime. The
distribution thus becomes superpoissonian at large $\kappa$. In
fact, in the limit $\kappa\to\infty$ one can derive the full
distribution function $P_{\kappa\to\infty}(\tau_m)$ using extreme
value statistics~\cite{Lubensky, ledoussal}:
\be
{\cal P}_{\kappa\to\infty}(\tau_m)\approx {\epsilon/ \kappa}\,
G(\tau_m\,\epsilon/\kappa)
\ee
with
\be
G(x)={1\over\sqrt{\pi x}}\,\mathrm e^{-x/4}-{1\over 2}{\rm
erfc}(\sqrt{x}/2),
\ee
where ${\rm erfc}(x)$ is the complimentary error function. It is
easy to check that this distribution indeed reproduces correct
expressions for the mean and the variance. We emphasize that while
the thermal fluctuations of the unzipping length become negligible
near the transition, the fluctuations due to different realizations
of point disorder are enhanced and lead to a wider-than-Poissonian
distribution of $\tau_m$.

\begin{figure}[h]
\center
\includegraphics[width=9cm]{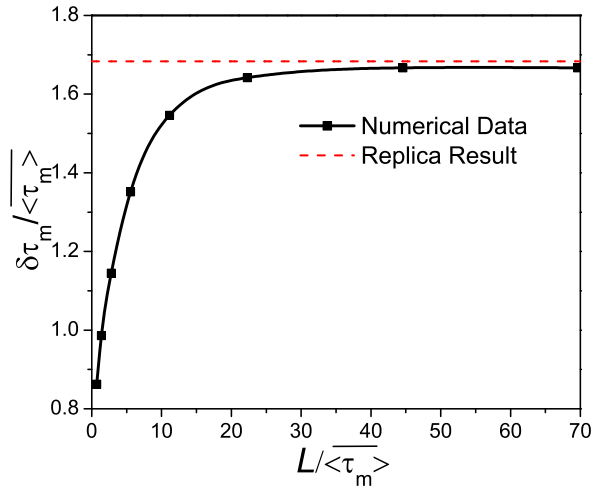}
\caption{Dependence of the relative width of the distribution
$\delta \tau_m/\overline{\langle \tau_m\rangle}$ on the system size.
Symbols correspond to the actual data, the solid line is the guide
to the eye, and the dashed line corresponds to the replica result in
the thermodynamic limit.}\label{fig_var}
\end{figure}

To check these results and uncover subtleties that might
arise in experiments, we performed direct numerical simulations of
the partition function of the free energy (\ref{fz}). For this
purpose we considered a discrete version of the problem where the
partition function is
\begin{equation}
Z=\prod_l \mathrm e^{-\epsilon m_l+\sum_{l^\prime=1}^l U(m_l)}.
\end{equation}
Here $U(m_l)$ is the random potential uniformly distributed in the
interval $[-U_0,U_0]$ so that the disorder variance is
$\Delta=\overline{U^2(m_l)}=U_0^2/3$. For the simulations we choose
$\epsilon=\ln(1.2)-0.18\approx 0.00232$ and $U_0=0.3$, which gives
$\Delta=0.03$, $\kappa\approx 6.46$ and according to both
Eq.~(\ref{zav}) and numerical simulations $\overline{\langle
\tau_m\rangle}\approx 2860$. Then we computed $\delta
\tau_m/\overline{\langle \tau_m\rangle}$ using both Eq.~(\ref{z2})
and performing numerical simulations. For the chosen parameters the
equation~(\ref{z2}) gives $\delta \tau_m/\overline{\langle
\tau_m\rangle}\approx 1.68$, while the numerical simulations  yield
$\delta \tau_m/\overline{\langle \tau_m\rangle}\approx 1.67$.
Clearly the results are very close to each other and the small
discrepancy can be attributed to the discretization error. In
Fig.~\ref{fig_var} we plot dependence of
$\delta\tau_m/\overline{\langle \tau_m\rangle}$ vs. system size.
It is obvious from the figure that in the thermodynamic limit
$L\to\infty$ the replica result is in excellent agreement with
numerical simulations. We mention that numerical simulations of
$\delta \tau_m$ show very strong finite size effects. Therefore one
has to go to very large $L\gtrsim 50 \overline{\langle
\tau_m\rangle}$ in order to approach the thermodynamic limit for the
width of the distribution.

\begin{figure}[h]
\center
\includegraphics[width=9cm]{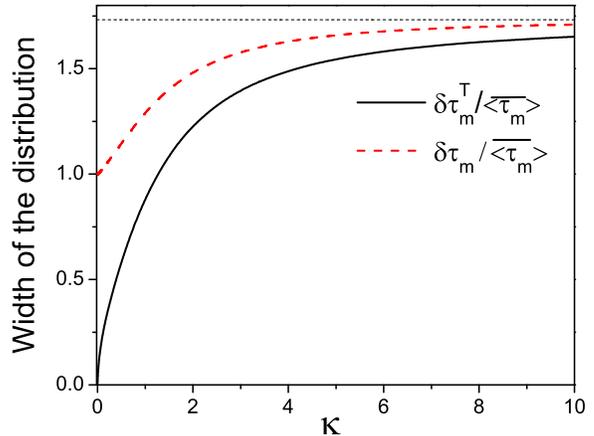}
\caption{Dependence of the relative width of the
distribution on the disorder parameter $\kappa$. The two curves
correspond to different averaging over temperature and disorder (see
text for details). The horizontal line at $\sqrt{3}$ denotes the
asymptotic value of both $\delta \tau_m$ and $\delta \tau_m^T$ at
$\kappa\to\infty$}\label{fig_dz}
\end{figure}

Depending on the system the quantity $\overline{\langle
\tau_m^2\rangle}$ is not always experimentally accessible. For
example, in the unzipping experiments it is easier to measure
thermal average, $\langle \tau_m\rangle$, in each experimental run.
We note that this quantity has sample to sample fluctuations only
due to the presence of disorder. Then the variance of the
distribution will be characterized by $\overline{\langle
\tau_m\rangle^2}$. The difference between the two expectation values
is given by $w$ found in Eq.~(\ref{fav}). Defining $(\delta
\tau_m^{T})^2=\overline{\langle \tau_m\rangle^2}-\overline{\langle
\tau_m\rangle}^{\,2}$ and using Eqs.~(\ref{z2}) and (\ref{fav}) we
find that $\delta \tau_m^T/\overline{\langle \tau_m\rangle}\approx
\sqrt{\kappa/2}$ in the weak disorder limit ($\kappa\ll 1$) and
$\delta \tau_m^T/\overline{\langle \tau_m\rangle}\approx
\sqrt{3}-1/(\sqrt{3}\kappa)$ in the opposite limit $\kappa\gg 1$. We
plot both $\delta \tau_m^T$ and $\delta \tau_m$ versus the disorder
parameter $\kappa$ in Fig.~\ref{fig_dz}.
The same issue of importance of the order of thermal and disorder
averaging appears in the calculation of the higher moments of
$\tau_m$, becoming irrelevant only in the limit ($\kappa\to\infty$),
which effectively corresponds to the zero temperature case.

Before concluding this section let us make a few remarks about the
rebinding transition, i.e., the rezipping that occurs with decreasing force. One can consider a similar setup with a lower
end of the flux line fixed at the bottom of the columnar pin and the
top end is pulled away from the pin with a force $f$. However, now
we will be interested in $f>f_c$. Then clearly most of the flux line
will be unzipped from the pin except for a portion near the bottom
end. If $f$ is very large, the length of the bound segment $\tilde
\tau_m$ near the sample boundary is small. However as $f$ decreases and approaches $f_c$ from
above, the length of this segment increases and finally diverges at
the transition. This rebinding transition can be described in a
similar spirit to the unbinding. For example instead of the free
energy (\ref{fz}) one has to deal with
\begin{equation}
\mathcal F(\tilde \tau_m)=|\epsilon|
\tilde\tau_m+\int_{0}^{\tilde\tau_m} d \tau_m' U(\tau_m').
\label{fzr}
\end{equation}
As we already noted the free energies (\ref{fz}) and (\ref{fzr}) are
equivalent up to an unimportant constant equal to the total disorder
potential of the pin: $\int_0^L d\tau_m' U(\tau_m')$. We conclude that the unbinding and rebinding
transitions for a single flux line on a disordered columnar pin are
identical. In other words, statistical properties of $\tau_m$ for a
given $f=f_c-\delta f$ are identical to those of $\tilde\tau_m$ for
$f=f_c+\delta f$.

\subsection{Unzipping from a planar defect without excursions.}
\label{replica_2D}

We now generalize the ideas of the previous section to the more
complicated problem of unzipping of a single flux line from a
disordered twin plane. As before we ignore excursions out of the
plane for the bound part of the flux line. Let us consider the
rebinding transition first. That is we assume that $f$ is slightly
greater than $f_c$ and we study the statistics of the
bound part of the flux line. We again assume that the flux line is
pinned at the bottom of the plane ($\tau=0$) and unbinds for $\tau$ larger than some
$\tilde\tau_m$.

The point disorder potential now depends on the two coordinates
$\tau$ and $z$ spanning the twin plane. Using Eq. (\ref{free_unzip}) the
partition function reads:
\begin{eqnarray}
&&Z=\int_0^L d\tilde\tau_m \int Dz(\tau^\prime)
\exp\biggl[-{f^2\over
2\gamma}\tilde\tau_m-V\tilde\tau_m\nonumber\\
&&~~~-\beta\int_0^{\tilde\tau_m} d\tau^\prime \left({\gamma\over
2}\left({dz\over d\tau^\prime}\right)^2+
\mu(\tau^\prime,z^\prime)\right)\Biggr],
\end{eqnarray}
where $V<0$ is the mean attractive potential of the twin plane and
we have dropped the unimportant $L$-dependent factors. As before, we
assume a Gaussian random noise with zero mean and
\begin{equation}
\overline{\mu(\tau_1,z_1)\mu(\tau_2,z_2)}=\sigma
\delta(\tau_1-\tau_2)\delta(z_1-z_2).
\end{equation}
We also introduce $\epsilon=-f^2/(2\gamma)-V$. Note that for the
rebinding transition $\epsilon<0$. After replicating the partition
function and averaging over point disorder we find
\begin{eqnarray}
&&\overline{Z^n}=n!\int\limits_{0}^L
d\tilde\tau_n\int\limits_{\tilde\tau_n}^{L}d\tilde\tau_{n-1}\ldots
\int\limits_{\tilde\tau_2}^L d\tilde\tau_1\int
Dz_1(\tau_1^\prime)\dots Dz_n(\tau_n^\prime)\nonumber\\
&&~~~~~~~\exp\left[\sum_{\alpha=1}^n
\epsilon\tilde\tau_\alpha+\!\!\!\int\limits_{\tilde\tau_{\alpha+1}}^{\tilde\tau_{\alpha}}
\!\!\!d\tau_\alpha^\prime \mathcal
L_{\alpha}[z_1(\tau_1^\prime),\ldots,
z_\alpha(\tau_\alpha^\prime)]\right],
\end{eqnarray}
where we define $\tilde\tau_{n+1}\equiv 0$ and $\mathcal L_\alpha$
is the Euclidean Lagrangian corresponding to the Hamiltonian
($\mathcal H_\alpha$) of $\alpha$ interacting particles~\cite{Kardar}:
\be
\mathcal H_\alpha=-{\sigma\over 2}\alpha-{1\over
2\gamma}\sum_{\beta=1}^{\alpha} {\partial^2\over\partial
z_\beta^2}-\sigma\sum_{1\leq\beta<\gamma\leq\alpha}
\delta(z_\beta-z_\gamma).
\ee
Close to the rebinding transition, we anticipate
$\tilde\tau_m\to\infty$ and thus the mean separation between the
rebinding times of different replicas $\tilde\tau_\alpha$ and
$\tilde\tau_{\alpha-1}$ diverges. Therefore the contribution to the
partition function coming from integration over $\tau_\alpha$ will
be dominated by the ground state of configurations with $\alpha$
replicas. In this case we can significantly simplify the partition
function and evaluate it analytically:
\begin{eqnarray}
&&\overline{Z^n}=n!\int\limits_{0}^L
d\tilde\tau_n\int\limits_{\tilde\tau_n}^{L}d\tilde\tau_{n-1}\ldots
\int\limits_{\tilde\tau_2}^L d\tilde\tau_1\\
&&\exp\left[\sum_{\alpha=1}^n  \epsilon\tilde\tau_\alpha+(\mathcal
E_{\alpha}-\mathcal E_{\alpha-1})
\tilde\tau_{\alpha})\right].\nonumber
\end{eqnarray}
Here $\mathcal E_\alpha$ is the ground state energy of $\mathcal
H_{\alpha}$ with a subtracted term linear in $\alpha$, that just
renormalizes $f_c$. Close to the transition $\epsilon$ is linear
in the difference $f-f_c$. The energy, $\mathcal E_\alpha$, was
computed in Ref.~[\onlinecite{Kardar}]:
\begin{equation}
\mathcal E_\alpha=-{\sigma^2\gamma\over
12}\alpha^3=-\xi\alpha^3.
\end{equation}
Upon integrating over $\tilde\tau_\alpha$ one obtains
\be
\overline{Z^n}=n!\prod_{\alpha=1}^n {1\over
|\epsilon|\alpha-\xi\alpha^3}\to\prod_{\alpha=1}^n {1\over
|\epsilon|-\xi\alpha^2}
\label{Z_n0}
\ee
The product above can be reexpressed in terms of $\Gamma$-functions,
which in turn allows for a straightforward analytic continuation to
$n\to 0$:
\be
\overline{Z^n} ={1\over
\xi^n}{1\over1+n{\sqrt{\xi}\over\sqrt{|\epsilon|}}}
{\Gamma\left({\sqrt{|\epsilon|}\over\sqrt{\xi}}-n\right)\over
\Gamma\left({\sqrt{|\epsilon|}\over\sqrt{\xi}}+n\right)}.
\label{Z_n1}
\ee
Using this expression we obtain the free energy and the mean
length of the localized segment:
\begin{equation}
\mathcal F=-\lim_{n\to 0}{\overline{Z^n}-1\over n}=\ln
\xi+{\sqrt{\xi}\over\sqrt{|\epsilon|}}+2\Psi\left({\sqrt{|\epsilon|}\over
\sqrt{\xi}}\right),
\label{f_2d}
\end{equation}
\begin{equation}
\overline{\langle \tilde\tau_m \rangle}={\partial \mathcal
F\over\partial |\epsilon|}=-{\sqrt{\xi}\over
2|\epsilon|^{3/2}}+{1\over
\sqrt{|\epsilon|\xi}}\Psi^{(1)}\left({\sqrt{|\epsilon|}\over
\sqrt{\xi}}\right)
\label{tau_2d}
\end{equation}
where as before $\Psi^{(n)}(x)$ stands for the $n$th derivative of
the digamma function. This expression has the asymptotic behaviors:
\begin{eqnarray}
&&\overline{\langle \tilde\tau_m \rangle}\to
{1\over\epsilon}\qquad\quad~ \xi\ll
|\epsilon|\nonumber\\
&&\overline{\langle \tilde\tau_m \rangle}\to {\sqrt{\xi}\over
2|\epsilon|^{3/2}} \quad \xi\gg|\epsilon|.
\end{eqnarray}
This scaling confirms the crossover between exponents $\nu=1$ and
$\nu=3/2$ for the rebinding transition to a two-dimensional
disordered plane predicted by the simple scaling argument leading to
Eq.~(\ref{valuenu}).

In a similar way one can also consider an unzipping transition with
$f \leq f_c$. One finds an expression for the partition function
which is identical to (\ref{Z_n0}) with the substitution
$\xi\to-\xi$. Note however, that the analytic continuation of the
product (\ref{Z_n1}) results in a complex partition function and
hence a complex free energy. It thus appears that the analytic
continuation of the product (\ref{Z_n0}) to noninteger values of $n$
is not unique. One can always multiply it by any periodic function
of $n$, which is equal to unity when the argument is integer. While
we were able to find some real-valued analytic continuations of
$\overline{Z^n}$ to negative values of $\xi$, these continuations
did not lead to physically sensible results.

Because of the ambiguity of the analytic continuation and some
approximations used to derive Eqs.~(\ref{f_2d}) and (\ref{tau_2d})
we also performed numerical simulations for the vortex unzipping
from a disordered twin plane.

For numerical simulations we are using the lattice version of the
model, where in each step along the $\tau$ direction the vortex can
either move to the left or the right one lattice spacing. Note that
because we neglect excursions the vortex motion occurs strictly
within the plane until the vortex is unbound. Then the restricted
partition function for the bound part of the flux line, $Z(x,\tau)$,
which sums over the weights of all path leading to $x,\tau$,
starting at $x=0,\tau=0$ satisfies the recursion
relation~\cite{Kardar}
\beq
&& Z(x,\tau+1)=e^{\mu(x,\tau+1)}\big[J Z(x-1,\tau) +J
Z(x+1,\tau)\nonumber\\
&&~~~~~~~~~~~~~~~~~~~~~~~~~~~~~~+(1-2J) Z(x,\tau)\big].
\label{eqz1}
\eeq
We assume that $\mu(x,\tau)$ is uniformly distributed in the
interval $[-U_0,U_0]$ implying as before the variance $\sigma=U_0^2/3$. The
variable $J$ controls the line tension. In the continuum limit $J\ll
1$ and $U_0\ll 1$ the equation (\ref{eqz1}) reduces to the
Schr\"odinger equation:
\be
{\partial Z\over \partial\tau}=-\mathcal H Z(x,\tau)
\label{Z_tauu}
\ee
with the Hamiltonian given by Eq.~(\ref{H}) with $\gamma=2J$ and
$f=0$ (there is no force acting on the flux line within the
plane). We note that even if the parameters of the discrete model are not
small we still expect that Eq.~(\ref{Z_tauu}) remains valid at long
length and time scales. However, the relation between the
microscopic parameters of the discrete model and the parameters of
the effective coarse-grained Hamiltonian (\ref{H}) is more
complicated.

In our simulations we evaluated numerically the free energy of the
bound part of the vortex line for each realization of point disorder
and used the analytical expression for the free energy of the
unbound part, for which point disorder can be neglected. The latter
is given by Eq.~(\ref{free_unzip}). This free energy is controlled
by a single parameter $f^2/(2\gamma)$. Use of the analytic result
(\ref{free_unzip}) significantly simplifies calculations of
$\overline{\langle \tau_m\rangle}$ and allows us to perform large
scale simulations.

First we verify the scaling (\ref{nu}) with $\nu=3/2$ at the
unzipping transition. To do this we perform standard finite size
scaling procedure.
\begin{figure}[ht]
\center
\includegraphics[width=9cm]{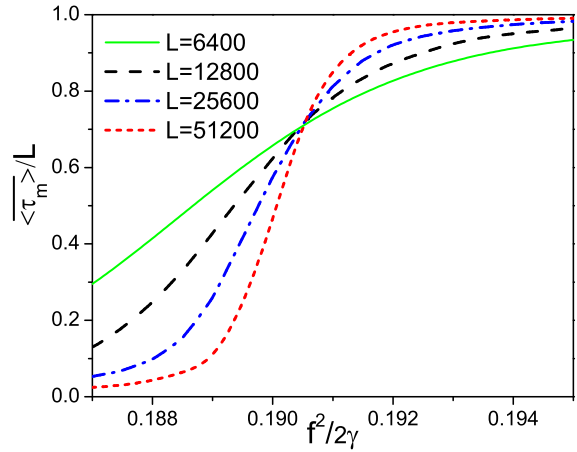}
\caption{Ratio of the unzipping length
$\overline{\langle\tau_m\rangle}$ to the system size $L$ as a
function of $f^2/2\gamma$ for different system sizes. Here $f$ is
the external force and $\gamma$ is the line tension of the vortex
(see Eqs.~(\ref{free_unzip1} and (\ref{free_unzip}))). According to
the scaling relation (\ref{scaling1}) the crossing point corresponds
to the unzipping transition. In simulations the parameters of the
microscopic model (\ref{eqz1}) are chosen to be $J=0.2$, $U_0=2$}
\label{fig3}
\end{figure}
In Fig.~\ref{fig3} we show dependence of the ratio
$\overline{\langle\tau_m\rangle}/L$ on the parameter $f^2/(2\gamma)$
for four different sizes. As we expect from the scaling relation
(\ref{scaling1}) the three curves intersect at the same point
corresponding to the unzipping transition ($g_0\approx 0.7$). Once
we determine the crossing point corresponding to the critical force
$f_c$ we can verify the scaling relation (\ref{scaling1}) with
$\nu=3/2$.
\begin{figure}[ht]
\center
\includegraphics[width=9cm]{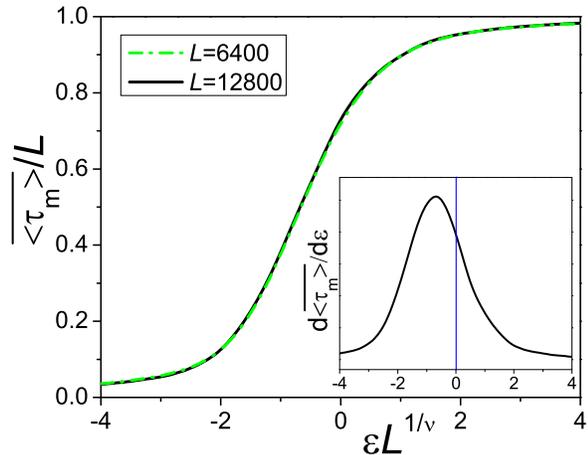}
\caption{Data collapse of $\overline{\langle\tau_m\rangle}/L$ as a
function of $\epsilon L^{1/\nu}$ with the exponent $\nu=3/2$ for two
different system sizes (see Eq.~(\ref{scaling1})). The parameters of
the model are the same as in Fig.~\ref{fig3}. The inset shows
derivative of $\overline{\langle\tau_m\rangle}$ with respect to
$\epsilon$ for $L=12800$. Clearly the scaling function is asymmetric
with respect to $\epsilon\to -\epsilon$. Thus the unbinding and
rebinding transitions are not equivalent.}
\label{fig:collapse2D}
\end{figure}
In Fig.~\ref{fig:collapse2D} we plot
$\overline{\langle\tau_m\rangle}/L$ versus the scaling parameter
$\epsilon L^{1/\nu}$ (see Eq.~(\ref{scaling1})) with $\nu=3/2$ for
two different system sizes. Clearly the data collapse is nearly
perfect, which proves the validity of the scaling (\ref{scaling})
with $\nu=3/2$ for the unzipping of a flux line from a twin plane.
The inset shows the derivative of $\overline{\langle\tau_m\rangle}$
with respect to $\epsilon$. Clearly this derivative is asymmetric
with respect to $\epsilon\to\ -\epsilon$, implying that there is no
symmetry between the unbinding and rebinding transitions. This is
contrary to the unzipping from a columnar pin with no excursions,
where such a symmetry does exist.

Next we turn to verifying the analytic prediction for
$\overline{\langle\tau_m\rangle}$, Eq.~(\ref{tau_2d}). As we argued
above the parameter $\zeta$ describing the disorder strength can be
easily extracted from microscopic parameters of the model only in
the continuum limit $U_0\gg 1$, $J\ll 1$. Unfortunately, it is not
possible to do simulations directly in the continuum limit ($J\ll 1$
and $U_0\ll 1$). Indeed as Eq.~(\ref{tau_2d}) suggests in order to
see the scaling exponent $\nu=3/2$ one needs to go to length scales much
larger than $1/\xi$, where $\xi=\sigma^2 J/12=U_0^4 J/36$. If
$J\ll 1$ and especially $U_0\ll 1$ then one has to simulate
extremely large system sizes where $L$ is larger than $10^7$ for
$U_0=0.1$ and $J=0.1$. Therefore we perform simulations in the regime
where $J$ and especially $U_0$ are appreciable. We then
regard $\xi$ as a fitting parameter of the model which should be
equal roughly to $U_0^4 J/36$.
\begin{figure}[ht]
\center
\includegraphics[width=9cm]{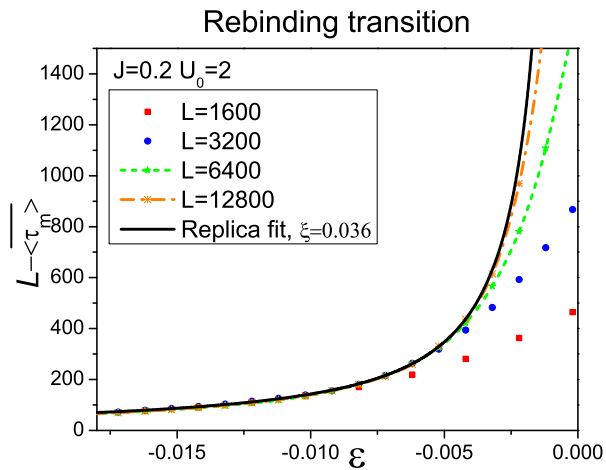}
\caption{Dependence of the length of the bound part of the flux line
to the twin plane $L-\overline{\langle\tau_m\rangle}$ on $\epsilon$
for the rebinding transition. Different curves correspond to
different system sizes. The solid black line is the best single
parameter fit using Eq.~(\ref{tau_2d}) with $\xi$ being the
fitting parameter.}
\label{fig:replica1}
\end{figure}
In Fig.~\ref{fig:replica1} we show results of numerical simulation
for the rezipping length $L-\overline{\langle\tau_m\rangle}$ on the
detuning parameter $\epsilon$ for different system sizes. The solid
black line is the best single-parameter fit to the data using the
analytic expression (\ref{tau_2d}). The fitting parameter $\xi$
found from simulations is $\xi \approx 0.036$, while a continuum
estimate $U_0^4 J/36$ gives $\xi \approx 0.089$, which is very
reasonable given that this estimate is valid only at $U_0\ll 1$. We
also performed similar simulations for $U_0=1.5$ and got a very good
fit with (\ref{tau_2d}) for $\xi=0.018$, while the continuum
estimate gives $\xi \approx 0.028$. We thus see that indeed as
$U_0$ decreases the fitting parameter $\xi$ becomes closer to the
continuum expression.

While we were not able to derive a closed analytic expression for
$\overline{\langle\tau_m\rangle}$ for the unbinding transition, we
performed numerical simulations. As the inset in
Fig.~\ref{fig:collapse2D} suggests the transition is highly
asymmetric. In fact this asymmetry persists in the thermodynamic
limit.
\begin{figure}[ht]
\center
\includegraphics[width=9cm]{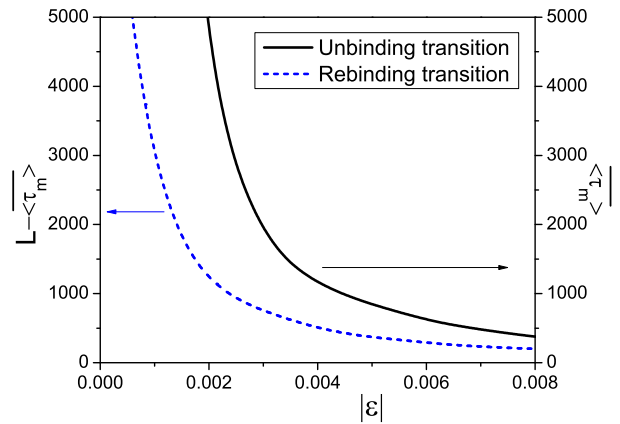}
\caption{Comparison of dependences of
$L-\overline{\langle\tau_m\rangle}$ for the rebinding transition and
$\overline{\langle\tau_m\rangle}$ for the unbinding transition on
$|\epsilon|$. We used the parameters of Fig.~\ref{fig3} with
$L=51200$. The finite size effects are negligible on the scale of
the graph. Both curves interpolate between $1/|\epsilon|$ dependence
at $|\epsilon|\gg \xi$ and $C/|\epsilon|^{3/2}$ at $|\epsilon|\ll
\xi$. However, the prefactor $C$ for the unbinding transition is
about three times larger than for the rebinding.}
\label{fig:unbind_rebind}
\end{figure}
In Fig.~\ref{fig:unbind_rebind} we plot
$L-\overline{\langle\tau_m\rangle}$ for the rebinding transition and
$\overline{\langle\tau_m\rangle}$ for the unbinding versus
$|\epsilon|$. Both curves interpolate between $1/|\epsilon|$
dependence at weak disorder $|\epsilon|\ll \xi$ and
$C/|\epsilon|^{3/2}$ dependence at strong disorder $|\epsilon|\gg
\xi$. However, the prefactor $C$ in front of $1/|\epsilon|^{3/2}$
is larger for the unzipping transition.

\subsection{Unzipping from a hard wall}
\label{Bethe_ansatz}

\begin{figure}[ht]
\centering
\includegraphics[scale=0.7]{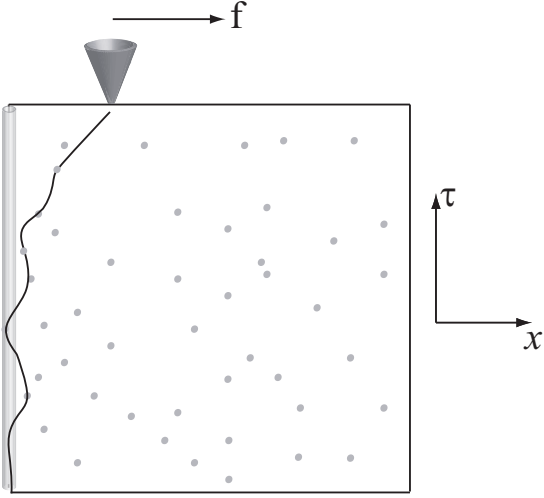}
\caption{\label{Bethe}  An illustration of the setup considered in
the Bethe Ansatz calculation. The flux line is restricted to the
half plane and a MFM tip is acting on it at the top of the sample.}
\end{figure}

As the next step we consider unzipping from  an attractive hard wall
in $d=1+1$ dimensions with point disorder in the bulk. Our method
is a straightforward generalization of the Bethe ansatz solution
found by Kardar in the absence of the external force~\cite{Kardar}.
The system is illustrated in Fig. \ref{Bethe}. Here the potential
experienced by the flux line, $V(x)$, has a short ranged attractive
part and an impenetrable core at $x=0$. While the scaling argument
is unchanged in this case, this problem has the merit of being
exactly solvable within the replica approach. Since most details of
the calculation are identical to those presented in
Ref.~[\onlinecite{Kardar}], here we only outline the solution.
After replicating the free energy Eq.~(\ref{f0dis}) along with the
contribution from the external field Eq.~(\ref{eq:unzipfe}) and
averaging over the disorder, the replicated sum over all path weights  connecting points $(0,0)$ and $(x,\tau)$,
$\overline{Z^n(x,t)}$, can be calculated from
\begin{equation}
\partial_\tau \overline{Z^n(x,t)} = -{\cal H} \overline{Z^n(x,t)} \;,
\end{equation}
with the initial condition $\overline{Z^n(x,0)}=\delta(x)$. The
replicated system describes $n$ {\it attractively interacting
bosons} with a non-Hermitian Hamiltonian ${\cal H}$ given by
\begin{eqnarray}
{\cal H}&=& \sum_{\alpha=1}^n \left[ -\frac{1}{2\gamma}
 \partial^2_{x_\alpha}-f
\partial_{x_\alpha}+ V(x_\alpha) \right]\nonumber
\\&-&\sigma \sum_{\alpha < \beta}
\delta(x_\alpha-x_\beta) -\frac{1}{2} \sigma n   \;.
\end{eqnarray}

In Ref.~[\onlinecite{Kardar}] the problem was solved for $f=0$ using
the Bethe Ansatz. The boundary conditions were that the ground state
wave function should vanish at large $x$ should decay as
$\exp(-\lambda x)$ for the particle closest to the wall.  One then
finds that for the permutation ${\bf P}$ of particles such that
$0<x_{P1}<x_{P2}< \ldots <x_{Pn}$ the wave function for $f=0$ is
\begin{equation}
\Psi_{f=0}\sim\exp \left(-\sum_{\alpha=1}^n \kappa_\alpha x_{P
\alpha}\right) \;.
\end{equation}
Here $\kappa_\alpha = \lambda+2(\alpha-1)\kappa$, $\kappa=
\sigma\gamma/2$. Taking the zero replica limit it was
found~\cite{Kardar} that for weak disorder ($\sigma\gamma/2 <
\lambda$) the vortex is bound to the wall while for strong
disorder ($\sigma\gamma/2 > \lambda$) it is unbound.

The ground state wave function for the {\it non-zero} value of the force
can be obtained by noting that the non-Hermitian term acts like an
imaginary vector potential. In particular, it can be gauged away when
the vortices are bound to the wall as discussed in Sec.
\ref{sectioncleancase} (see Eqs.~(\ref{eq:gauge1}) and
(\ref{eq:gauge2})). This imaginary gauge transformation gives
\begin{equation}
\Psi_{f}=\Psi_{f=0}\exp\left( \sum_{\alpha=1}^n fx_\alpha \right)
\;,
\end{equation}
which implies that the solution is
\begin{equation}
\Psi_{f}=\exp \left(-\sum_{\alpha=1}^n \tilde{\kappa}_\alpha x_{P
\alpha}\right) \;,
\end{equation}
with $\tilde{\kappa}_\alpha = \lambda+2(\alpha-1)\kappa-f$. The
effect of the force is simply to shift all the $\kappa_\alpha$'s by a
constant. The average localization length (which satisfies near the
transition $\langle x _m\rangle \simeq f_c \langle \tau_m \rangle /
\gamma$) is then given by
\begin{equation}
\langle x_m \rangle={1\over \tilde Z_n n}\int_0^\infty
\prod_{j=1}^n dx_j \left[\sum_{j=1}^n x_j\right]\,\Psi_f(x_j),
\label{eq:Kardarresult}
\end{equation}
where $\tilde Z_n=\int_0^\infty  \prod_{j=1}^n dx_j \Psi_f(x_j)$.
Note that the normalization factor $\tilde Z_n$ in the equation
above is formally equivalent to the partition function (\ref{tuam2})
for the unzipping from a columnar pin without excursions if we
identify $\lambda-\kappa-f$ with $\epsilon$ and $\kappa$ with
$\Delta/2$. This equivalence implies that $\langle x_m\rangle$ for
the unzipping from a hard wall has the same statistical properties
as $\overline{\langle\tau_m\rangle}$ for the unbinding from a
columnar pin (for more details see Ref.~[\onlinecite{kp}]). In particular, the
unzipping problem  has a crossover from $\langle x_m\rangle \sim
1/(f_c-f)$ for $\lambda-f\gg\kappa$ to $\langle x_m\rangle \sim 1/
(f_c-f)^{3/2}$ in the opposite limit.

This example confirms another prediction of
the simple scaling argument: the critical exponents for the
unbinding transition are determined only by the dimensionality of
the defect even if the disorder is also present in the bulk of the system.

\subsection{Unzipping from a columnar pin with excursions into the bulk.}
\label{sec:numerics}

\begin{figure}[ht]
\centering
\includegraphics[width=9cm]{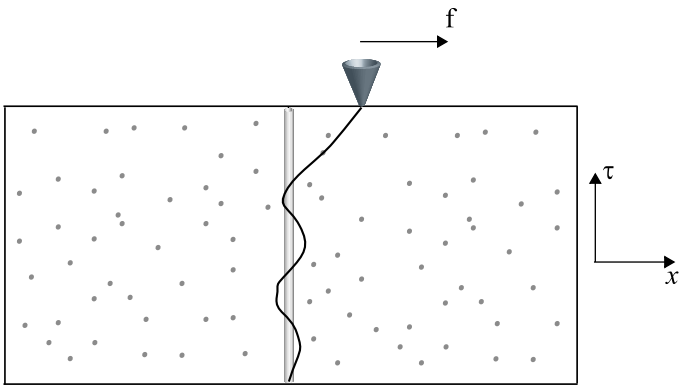}
\caption{A setup illustrating unzipping from a
columnar pin in $d=1+1$ dimensions with excursions into the bulk.}
\label{fig:unzip_1D}
\end{figure}

In this section we consider the setup similar to Sec.~\ref{replica},
namely unzipping from a columnar defect in $d=1+1$ dimensions, but
allowing excursions of the flux line to the bulk (see
Fig.~\ref{fig:unzip_1D}).
Unfortunately there is no analytic solution available for this
problem. Therefore we present only numerical results. As in
Sec.~\ref{replica_2D} we consider a lattice version of the model
where in each step along the $\tau$ direction the vortex can either
move to the left or the right one lattice spacing. The attractive
potential was placed at $x=0$. The restricted partition function of
this model, $Z(x,\tau)$, which sums over the weights of all path
leading to $x,\tau$, starting at $x=0,\tau=0$ satisfies the
recursion relation~\cite{Kardar}:
\begin{eqnarray}
&&Z(x,\tau+1)=\delta_{x,0}(e^{V}-1)Z(0,\tau) \nonumber\\
&&+e^{\mu(x,\tau+1)}\left[J e^f Z(x-1,\tau) +J
e^{-f}Z(x+1,\tau)\right]. \label{eqz}
\end{eqnarray}
Similarly to Eq.~(\ref{eqz1}) we assume that $\mu(x,\tau)$ is
uniformly distributed in the interval $[-U_0,U_0]$ implying the
variance $\sigma=U_0^2/3$. The variable $J$ controls the line
tension, $V$ is the attractive pinning potential, and $f$ is proportional
to the external force. In the continuum limit $J\ll 1$, $f\ll 1$,
and $U_0\ll 1$, equation (\ref{eqz}) reduces to the Schr\"odinger
equation:
\be
{\partial Z\over \partial\tau}=-\mathcal H Z(x,\tau)
\label{Z_tau}
\ee
with the Hamiltonian given by Eq.~(\ref{H}) with $\gamma=2J$.

For the simulations we have chosen particular values of $J=0.1$ and
$V=0.1$. As before we work in units such that $k_B T=1$. In the
results described below the partition function was evaluated for
each variance of the disorder for several systems of finite width
$w=2L_x$ averaging over the time-like direction (typically $\tau
\simeq 10^6$ ``time'' steps) with the initial condition $Z(0,0)=1$
and $Z(x,0)=0$ for $x \neq 0$.

\begin{figure}
\center
\includegraphics[width=8.5cm]{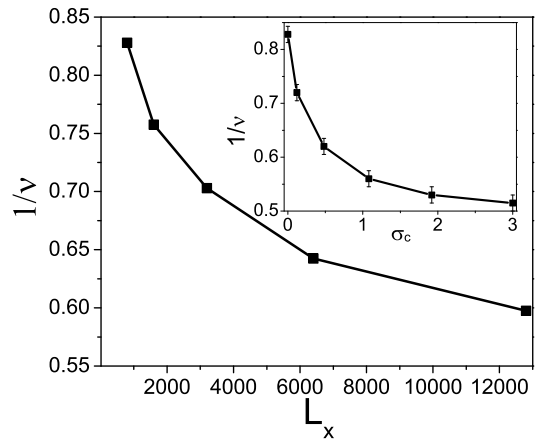}
\caption{Effective exponent $1/\nu$ versus $L_x$ for a fixed
strength of point disorder $\sigma=0.03$. The results are
consistent with the general argument that this exponent should
saturate at $\nu=2$ as $L_x\to\infty$. The inset shows the same
exponent vs $\sigma_c$, the variance of additional point disorder
placed directly on the columnar pin extracted from two system
sizes $L_x=600$ and $L_x=1200$. It appears that $\nu\to 2$ as
$\sigma_c$ increases.} \label{fig1}
\end{figure}

To analyze the numerics we performed a finite size scaling analysis. In the spirit of Eq.~(\ref{nu}), in the vicinity of the transition we
expect the scaling form (compare Eq.~(\ref{scaling1})):
\begin{equation}
\overline{\langle\tau_m\rangle}=L_x \Phi\left[L_x(f_c-f)^\nu\right],
\label{scaling}
\end{equation}
where $\Phi$ is some scaling function.
Based on the results of previous sections we anticipate a smooth
interpolation between scaling exponents $\nu=1$ and $\nu=2$ with either
increasing $L_x$ or increasing strength of disorder at fixed $L_x$.
To perform the finite size scaling we obtain for each value of $L_x$
a value for the exponent $\nu$ from the best collapse of the
numerical data of two systems sizes $L_x$ and $L_x/2$. In
Fig.~\ref{fig1} we plot $1/\nu$ as a function of the system size
$L_x$. As can be seen the data is consistent with $\nu$ saturating
at $\nu=2$ for large systems. The crossover to $\nu=2$ is much more
rapid if the point disorder is enhanced near the columnar pin (see
the inset in Fig.~\ref{fig1}), as might be expected for damage
tracks created by heavy ion radiation.

Next, we test the behavior of the critical force as the
disorder strength is increased. According to our discussion in Sec.
\ref{scalingphasediagram}, we anticipate that in the absence of an
external force the flux line is always bound to the pin in $1+1$
dimensions. This is in contrast with the problem of unzipping from
the wall discussed in the previous section, where there is a
critical strength of the disorder, $\sigma_c$, which leads to an
unbinding transition for $f=0$. Note that the existence of a critical value
of the disorder is a direct consequence (see discussion in Sec.
\ref{scalingphasediagram}) of the excursions of the vortex from the
defect which, as argued above, do not modify the critical behavior
of the unzipping transition. The existence of a critical value of
the disorder is therefore strongly dependent on the dimensionality
of the problem.

\begin{figure}[ht]
\hspace{0.5cm}
\includegraphics[bb=1cm 1cm 20cm 25cm, scale=0.38, angle=90]{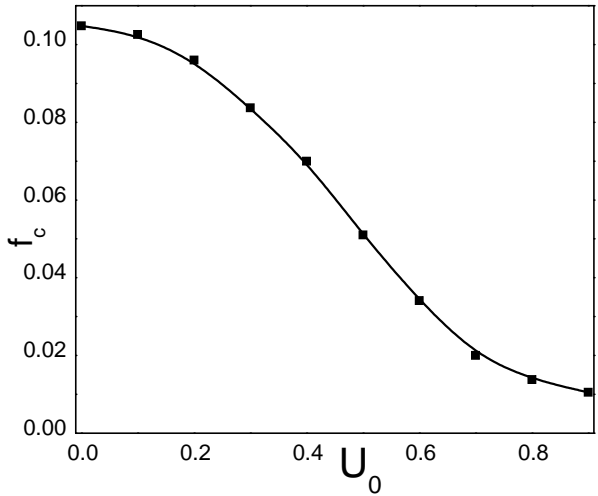}
\caption{Critical force for unzipping from a columnar defect in
$1+1$ dimensions as a function of the disorder
strength.}\label{fig8}
\end{figure}

In numerical simulations for each strength of disorder we determine
the critical force plotting the ratio
$\overline{\langle\tau_m\rangle}/L_x$ for two different sizes $L_x$
and using the scaling relation~(\ref{scaling}). Note that this ratio
does not depend on $L_x$ at $f=f_c$ (see also the discussion in
Sec.~\ref{replica_2D}). We checked that this is indeed the case.
Upon repeating this procedure for different disorder strengths we
obtain the dependence $f_c(U_0)$ which is plotted in
Fig.~\ref{fig8}.
The graph suggests that there is no unbinding transition at zero
tilt at any strength of disorder consistent with the scaling
argument presented in Sec.~\ref{scalingphasediagram} and those of
Ref.~[\onlinecite{HwaNatter}]. We point out that the strongest disorder
shown in the graph $U_0=0.9$ required samples quite extended in the
time-like direction, $L_\tau\approx 10^8$.

\section{Unzipping a Luttinger liquid}
\label{sec:Lutunzip}

\begin{figure}[ht]
\includegraphics[scale=0.6]{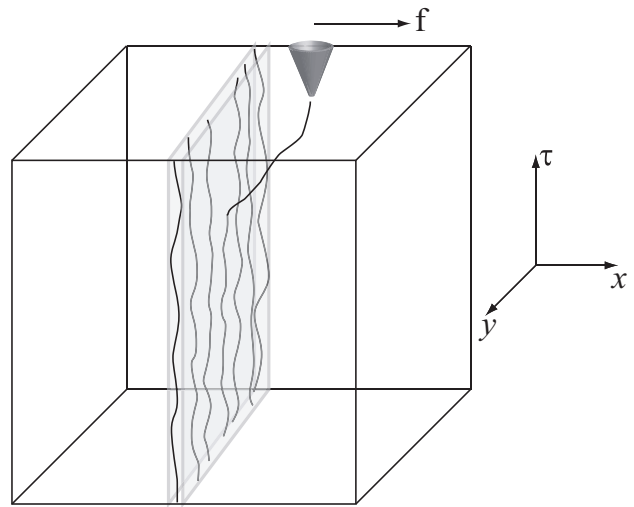}
\caption{Possible experimental setup for studying unzipping from
Luttinger Liquid. A MFM is used to pull a single vortex out of a
plane where the vortices are confined. The measured quantity is the
distance of the pulled vortex from the confining plane as a function
of the force $f$. }\label{fig9}
\end{figure}

We now turn to consider the effect of interactions on the unzipping
of single vortices. To do this we study a system where the vortices
are preferentially bound to a thin two-dimensional slab which is
embedded in a three-dimensional sample so that the density of
vortices in the slab is much higher than in the bulk.
Experimentally, this setup could be achieved using, for example, a
twin plane in YBCO or by inserting a thin plane with a reduced lower
critical field $H_{c1}$ (with, for example, molecular beam epitaxy)
into a bulk superconductor. The scenario we analyze is one where a
MFM is used to pull a single vortex out of the two-dimensional slab
(see Fig. \ref{fig9}). The physics of the vortices confined to two
dimensions is well understood and is analogous to a spinless Luttinger
liquid of bosons (see, e.g. Ref.~[\onlinecite{AHNS}]).

As we show below the dependence of the displacement of the vortex
from the two-dimensional slab on the force exerted by the MFM
depends on the physics of the two-dimensional vortex liquid which
resides in the slab. Specifically, the critical properties of the
unbinding transition depend on the ``Luttinger liquid parameter''
which controls the large-distance behavior of the vortex liquid. The
experimental setup can thus be used to probe the two-dimensional
physics of the vortices in the slab.

\subsection{Two-dimensional vortex liquids}

The physics of vortices in two dimensions is very well understood.
The vortices form a one-dimensional array located at position
$x_i(\tau)$. The density profile of the vortices is then given by
\begin{equation}
    n(x,\tau)=\sum_j \delta \left[ x-x_j(\tau)\right] \;,
\end{equation}
where $x$ and $\tau$ denote transverse and longitudinal coordinates
with respect to the vortices and $i$ is an index labeling the
vortices. By changing variable into the phonon displacement field
$u_j$ through $x_j(\tau)=a\left[j+u_j(\tau)\right]$, where $a$ is
the mean distance between vortex lines the free-energy of a
particular configuration can be written as:
\begin{equation}
    {\cal F}_0=\frac{a^2}{2} \int dx d\tau \left[ c_{11}
    (\partial_x u)^2 + c_{44} (\partial_\tau u)^2\right] \;.
\end{equation}
Here $c_{11}$ and $c_{44}$ are the compressional and the tilt moduli
respectively. After rescaling the variables $x$ and $\tau$ according
to
\begin{equation}
    x \to x \left(\frac{c_{11}}{c_{44}}\right)^{1/4} \;\; ,
    \; \tau \to \tau \left(\frac{c_{44}}{c_{11}}\right)^{1/4} \;,
\end{equation}
the free energy takes the isotropic form
\begin{equation}
    {\cal F}_0=\frac{A}{2}\int dx d\tau
    \left[ (\partial_x u)^2 + (\partial_\tau u)^2\right]
\end{equation}
with $A=a^2\sqrt{c_{11}c_{44}}$. The partition function is then
given by the functional integral
\begin{equation}
    Z=\int D u(x,\tau) e^{-S} \;,
\end{equation}
with $S=S_0={\cal F}_0/T$. In the limit of large sample sizes in the
``timelike'' direction  one can regard $Z$ as the zero temperature
partition function of interacting bosons~\cite{AHNS}. In this
language the imaginary time action can be written as
\begin{equation}
    S_0=\frac{\pi}{2g}\int dx d\tau
    \left[ (\partial_x u)^2 + (\partial_\tau u)^2\right] \;.
    \label{freeaction}
\end{equation}
Here we set $\hbar=1$ and identified the Luttinger-liquid parameter,
$g$, as
\begin{equation}
    g=\frac{\pi T}{A} \;.
    \label{Lutpara}
\end{equation}
The Luttinger-liquid parameter controls the long-distance properties
of the model. For vortices $g$ it is a
dimensionless combination of the compressional and tilt moduli, the
density of vortices and temperature.

Various properties of Luttinger liquids are well understood. For
example, the correlation function for the density fluctuations
$\delta n(x,\tau)=n(x,\tau)-n_0$, where $n_0=1/a$ is the mean
density, obeys
\begin{equation}
    \langle \delta n(x,\tau) \delta n(0,0) \rangle \simeq
    \frac{\cos\left( 2 \pi n_0 x \right)}{(x^2+\tau^2)^g} \;.
\end{equation}
There is quasi long-range order in the system and the envelope of
the density correlation function decays as a power law with the exponent
depending only on $g$. As we show below, $g$ can be probed by
unzipping a single vortex out of a plane which contains a $(1+1)$-dimensional vortex liquid.

In what follows we also consider the case where there is point
disorder present in the sample. The behavior will be strongly influenced by the behavior of the vortices
in two dimensions in the presence of disorder. This problem has been
studied in some detail in the past (see e.g. Ref.~[\onlinecite{pkn}] and
references therein). Here we briefly review features which will be
important in analyzing the unzipping problem. The most relevant (in
the renormalization group sense) contributions to the action from
the point disorder is
\begin{equation}
    S_{PD}=2\int dx d\tau R(x,\tau)
    \cos \left[2 \pi u(x,\tau) +\beta(x,\tau) \right] \;,
\end{equation}
where positive (negative) $R$ implies a repulsive (attractive)
potential between the vortices and the quenched random disorder. We assume, for simplicity, that $\beta(x,\tau)$ is
distributed uniformly between $0$ and $2 \pi$ and $R(x,\tau)$ has a
an uncorrelated Gaussian distribution with the variance $\Delta_0$:
\begin{equation}
    \overline{R(x_1,\tau_1)R(x_2,\tau_2)}=
    \Delta_0 \delta(x_1-x_2)\delta(\tau_1-\tau_2) \;,
\end{equation}
where the overbar, as before, represents averaging over disorder.

To analyze the disordered problem, similar to the single vortex case,
we use the replica trick. Then the replicated noninteracting part of
the action becomes
\begin{equation}
    S_0=\frac{\pi}{2g} \sum_{\alpha,\beta} \int
    \int dx d\tau \left[ \frac{\partial u_\alpha}{\partial \tau}
    \frac{\partial u_\beta}{\partial \tau} +\frac{\partial u_\alpha}{\partial x}
    \frac{\partial u_\beta}{\partial x}  \right] \left[ \delta_{\alpha,\beta}
    - \frac{\kappa}{g}  \right] \;.
\end{equation}
Here $u_\alpha(x,\tau)$ is the replicated phonon field and $\kappa$
is an off-diagonal coupling which is zero in the bare model but is
generated by the disorder. It plays the role of a quenched random
``chemical potential'' which is coupled to the first derivative of
the phonon field $u$. The replica indices, $\alpha$ and $\beta$ run
from $1$ to $n$ and at the end of the calculation one takes the
limit $n \to 0$. After replication the contribution from the point
disorder becomes
\begin{equation}
    S_{PD}=-\Delta_0 \sum_{\alpha,\beta} \int \int dx d\tau
    \cos 2 \pi \left[ u_\alpha (x,\tau) - u_\beta (x,\tau)  \right] \;.
\end{equation}
The combined action can be treated within the renormalization group
using a perturbation series near $g=1$ where a phase transition
between a vortex liquid and a vortex glass
occurs~\cite{fisher_v_glass}. By continuously eliminating degrees of
freedom depending on frequency and momentum within the shell
$\Lambda - \delta \Lambda < \sqrt{\omega^2+q^2} < \Lambda$, one
obtains the following renormalization group equations~\cite{Cardy,
pkn}
\begin{eqnarray}
    \frac{dg}{dl}&=&0 \\
    \frac{d \Delta}{dl}&=&2(1-g) \Delta - 2 C \Delta^2 \\
    \frac{d \kappa}{dl}&=&C^2 \Delta^2
\end{eqnarray}
Here $l$ is the flow parameter $\Lambda(l)=\Lambda e^{-l}$.  $C$ is
a non-universal constant which depends on the cutoff $\Lambda$. The
equations are subject to the initial conditions $\kappa(l=0)=0$ and
$\Delta(l=0)=\Delta_0$. Note that the Luttinger liquid parameter is
not renormalized.  Analyzing the flow equations it has been shown
that in the vortex liquid phase ($g>1$) the correlation of the
density fluctuation behaves in the vortex liquid phase as
\begin{equation}
    \langle \delta n(x,\tau) \delta n(0,0) \rangle
    \simeq \frac{1}{(x^2+\tau^2)^{g+\tilde{\kappa}/2}} \;,
\end{equation}
where $\tilde{\kappa}$ is a nonuniversal exponent. In the glass
phase ($g<1$) correlations decay faster than a power law, with
\begin{equation}
    \langle \delta n(x,\tau) \delta n(0,0) \rangle
    \simeq \exp \left( -(1-g)^2 \ln^2 \sqrt{x^2+\tau^2}\right) \;.
\end{equation}

In what follows we consider a setup in which a two dimensional array
of vortices, whose properties have been described above, is embedded
in a three dimensional bulk sample. As shown below when a single vortex
is unzipped into the bulk in a clean sample the critical properties
of the unzipping transition yield information on the properties of
the two dimensional vortex liquid. In particular, they provide a
direct measure of the Luttinger-liquid parameter. In the same setup
in a disordered sample we will show that the critical properties of
the unzipping transition will be modified. In particular, they can
yield information on the on the three-dimension wandering exponent
of a single vortex in a disordered sample.

\subsection{Unzipping a Luttinger liquid: The clean case}
Consider first an experiment where an attractive two-dimensional
potential holds vortices confined to it. A MFM then pulls a {\it
single} vortex out of the plane (see Fig. \ref{fig9}). We assume
throughout that the density of vortices in the three dimensional bulk
is so small that we can neglect interactions between the vortex that
is pulled out of the sample and vortices in the three dimensional
bulk. In this subsection only the clean case (no point disorder) will be studied.

We assume the MFM exerts a force ${\bf f}=f \hat{x}$. As in the
unzipping experiments discussed above we expect that for large
forces $f>f_c$ the vortex will be completely pulled out of the two
dimensional slab. Similar to the case of the unzipping of a single
vortex we write the free energy of the vortex as a sum of two
contributions. The first, ${\cal F}_u(\tau_m)$, arises from the part
of the vortex that is outside the two dimensional slab. The second
${\cal F}_b(\tau_m)$ is the change in the free-energy of the
vortices that remain inside the two dimension slab. As before
$\tau_m$ is the length along the $\tau$ direction which is unbound
from the two-dimensional slab. The free-energy of the unzipped part
is clearly identical to that calculated in Eq.~\ref{free_unzip} or
explicitly
\begin{equation}
    {\cal F}_u(\tau_m)= -  f^2 \tau_m/ 2\gamma \;.
    \label{eq:unzupfeagain}
\end{equation}

The calculation of the free-energy, ${\cal F}_b(\tau_m)$, is
somewhat more involved. Clearly there is a linear contributions due
to the length $\tau_m$ removed from the attractive potential of the
slab. However, in addition there is an extra contribution from the
energy of the dislocation, ${\cal F}_d(\tau_m)$, (see Fig.
\ref{fig9}) created in the two dimensional vortex array. This
contribution to the free-energy, as we show below, is {\it
non-linear} and controlled by the Luttinger liquid parameter $g$.
This non-linearity results, near the unzipping transition, in a
sensitivity of the critical properties to the value of $g$.

We leave the details of the calculation of the dislocation energy to
Appendix~\ref{App:dislocation} and present here only the key steps
of derivation.

In order to satisfy boundary conditions near the interface one can
use the method of images (see Fig.~(\ref{fig11})).  The free energy
of this dislocation pair can be calculated by standard methods (see
details in Appendix~\ref{App:dislocation}). In particular, at large
$\tau_m$ it behaves logarithmically (see e.g. Ref.~[\onlinecite{chakin}]):
\begin{equation}
    {\cal F}_d=\frac{T}{4g} \ln(\tau_m/a_0),
    \label{free_en_dis}
\end{equation}
where $a_0$ is the short range cutoff of the order of the distance
between flux lines. We note that the free energy of the dislocation
near the interface (\ref{free_en_dis}) is one half of the free
energy of a dislocation pair.

With the energy of the dislocation in hand we can now analyze the
properties of the unzipped length near the transition using the
methods used for analyzing the single vortex unzipping experiments.
The contributions to the free energy are from the unzipped part of
the vortex and the energy of the dislocation. Collecting all the
relevant terms, near the transition the free energy is given by
\be
  {\cal F}(\tau_m)={\cal F}_u(\tau_m)+{\cal F}_b(\tau_m)=
  \epsilon\tau_m+\frac{T}{4g}\ln(\tau_m/a_0)\;.
\ee 
The probability of finding a certain value of $\tau_m$ is then given
by
\begin{equation}
    P(\tau_m) \propto e^{-F(\tau_m)/T}=\frac{C}{\tau_m^{1/(4g)}}e^{-\epsilon\tau_m}.,
\end{equation}
where $C$ is the normalization constant. At the transition
$\epsilon=0$ the distribution becomes a pure power law in $\tau_m$.
Therefore, the average value of $\tau_m$ is very sensitive to the
value of $g$. In particular, for $g>1/4$ (i.e. for weakly
interacting flux lines) the behavior of $\langle\tau_m\rangle$ near
the transition is identical to that of a single vortex in the
absence of interactions with other vortices
\begin{equation}
    \langle \tau_m \rangle \sim {1\over\epsilon} \;.
\end{equation}
In contrast, for $1/8 < g < 1/4$ (stronger interactions) there is a
continuously varying exponent governing the transition
\begin{equation}
\langle\tau_m\rangle\sim {1\over \epsilon^{2-1/4g}} \;.
\end{equation}
And finally, for $g<1/8$ (strongly interacting flux lines) we find
that $\langle \tau_m\rangle$ does not diverge near the transition.
Note that even though in this regime the mean displacement remains
constant at the transition the higher moments of $\tau_m$ diverge
and are thus sensitive to $\epsilon$. The reason for this is at the
transition the distribution of $\tau_m$ is a power law.

\subsection{Unzipping from a twin plane with point disorder}
\label{3c}

We now consider the problem of unzipping a vortex from a
plane with many vortices in the presence of disorder. In the spirit
of the treatments presented in this paper, one needs to calculate the
free-energy of the unzipped part of the vortex ${\cal F}_u(\tau_m)$,
the free-energy of the bound part of the vortex ${\cal F}_b(\tau_m)$
and the {\it fluctuations} in both quantities averaged over
realizations of disorder. This can be done perturbatively near $g=1$. We again relegate details of the derivation of
the dislocation energy to
Appendix~\ref{App:dislocation1}. One conclusion from our
calculations is that the mean free energy of the dislocation near
the boundary is not affected by the disorder and is given by
Eq.~(\ref{free_en_dis}). Another important conclusion is that the
fluctuations of the free energy also depend logarithmically on
$\tau_m$:
\begin{equation}
    \overline{\delta {\cal F}^2_d(\tau_m)}= T^2\frac{\kappa(\infty)}{8g^2} \ln(\tau_m/a_0)
\end{equation}
for $g>1$ and
\begin{equation}
    \overline{\delta {\cal F}^2_d(\tau_m)}=T^2\frac{(1-g)^2}{4} \ln^2(\tau_m/a_0)
\end{equation}
for $g<1$.
\begin{figure}[ht]
\includegraphics[scale=0.6]{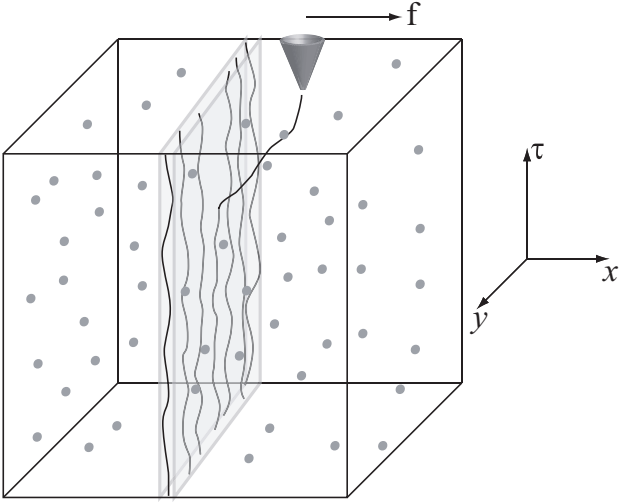}
\caption{Possible experimental setup for studying unzipping from
Luttinger Liquid in the presence of disorder.}\label{fig10}
\end{figure}
We note that in the case of many flux lines there is a weak
logarithmic dependence of free energy fluctuations on $\tau_m$ as
opposed to strong power law dependence in the case of a single flux
line (compare Eq.~(\ref{eq:fefluct})). This somewhat surprising
result is a consequence of the screening of strong power-law
fluctuations by other flux lines. We note that if the pinning of
flux lines by disorder is extremely strong so that tearing a single
flux line does not affect positions of other lines in the duration
of experiment, we are back to the single flux line physics and
$\overline{\delta \mathcal F_d^2}\propto \tau_m$.

To complete the analysis, we need to consider the free-energy
contribution from the unzipped part. Of particular of importance are
the free-energy fluctuations due to the disorder in the bulk of the
sample. As discussed in Sec. \ref{Sec2}, in a three dimensional
sample these grow as $\delta {\cal F}_u \propto m^{\omega(3)}$ with
$\omega(3) \simeq 0.22$. This contribution grows much quicker than
the contribution from the fluctuations in the free-energy of the
dislocation. Therefore following the ideas of Sec. \ref{Sec2} the
total free-energy is given by
\begin{equation}
{\cal F}(m)=a(f_c-f)\tau_m -b\tau_m^{\omega(3)}\;.
\label{fplane}
\end{equation}
where $a$ and $b$ are positive constants. Minimizing Eq. (\ref{fplane}) gives for the critical properties in
this case
\begin{equation}
    s \sim \frac{1}{(f_c-f)^{1.28}} \;.
\end{equation}
Thus screening disorder fluctuations in the plain by other flux
lines effectively enhances the role of disorder in the bulk. As the
result these unzipping experiments can serve as a probe of the
three-dimensional anomalous wandering exponent.

\acknowledgements

YK was supported by the Israel Science Foundation and thanks the
Boston University visitors program for hospitality. YK and DRN were
supported by the Israel-US Binational Science Foundation. Research
by DRN was also supported by the National Science Foundation,
through grant DMR 0231631 and through the Harvard Materials Research
Science and Engineering center through grant DMR0213805. AP was
supported by AFOSR YIP.

\appendix

\section{Calculation of the dislocation energy near the interface.}
\label{App:dislocation}

\begin{figure}[ht]
\includegraphics[scale=0.6]{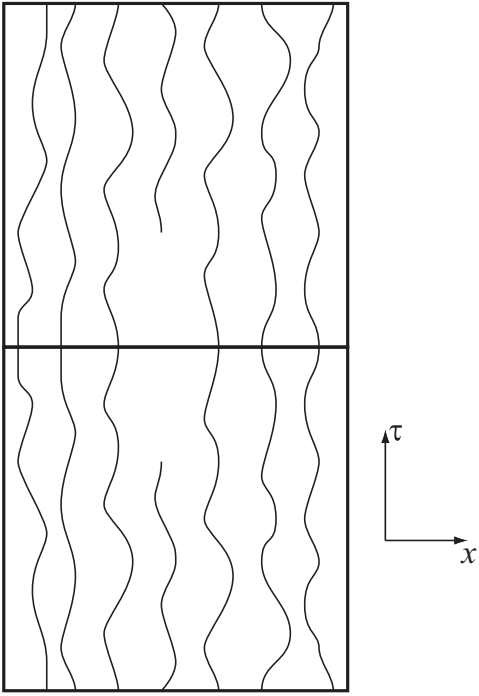}
\caption{An illustration of configurations of the phonon field considered by the method of images.}\label{fig11}
\end{figure}

To calculate the energy of the dislocation created by unzipping in Fig.~\ref{fig9}, standard methods can be
used with a slight modification. The need for a modification is due
to the requirement that vortex lines must exit normal to the
interface of the superconductor at the top of the sample which
ensures that the supercurrents are confined to the sample. One then
has to sum over phonon field configuration, $u_d(x,\tau)$, which
satisfy
\begin{equation}
    \left. \frac{\partial u_d(x,\tau)}{\partial \tau} \right \vert_{\tau=0}=0 \;.
\end{equation}
Here, we have chosen the upper boundary of the two-dimensional slab to be
located at $\tau=0$. The calculation can then be done using the
method of images, depicted in Fig. \ref{fig11}, by writing the
constrained field $u_d(x,\tau)$ in terms of an unconstrained field
$u(x,\tau)$
\begin{equation}
    u_d(x,\tau)=u(x,\tau)+u(x,-\tau) \;,
\end{equation}
and summing over all configurations of the unconstrained field. In
terms of the constrainted field the action is
\begin{equation}
    S=\frac{\pi}{2g}\int_{-\infty}^0 d \tau \int_{-\infty}^{\infty} dx
    \left[ (\partial_x u_d(x,\tau))^2+(\partial_\tau u_d(x,\tau))^2\right] \;,
\end{equation}
which rewritten in terms of the unconstrained field becomes
\begin{eqnarray}
    S&=&\frac{\pi}{2g}\int_{-\infty}^\infty d \tau \int_{-\infty}^{\infty} dx
    \left[ (\partial_x u(x,\tau))^2+(\partial_\tau u(x,\tau))^2 \right. \nonumber \\
    &+&\left.  \partial_x u(x,\tau)\partial_x u(x,-\tau)+\partial_\tau u(x,\tau)\partial_\tau u(x,-\tau) \right] \;.
\end{eqnarray}
In terms of the partial Fourier transform
\begin{equation}
    u(x,\tau)=\frac{1}{(2\pi)^{1/2}} \int d\omega e^{i \omega \tau} u(x,\omega) \;,
\end{equation}
the action becomes
\begin{equation}
    S=\frac{\pi}{2g} \int_{-\infty}^\infty d \omega \int_{-\infty}^{\infty}
    dx\left[ (\partial_x u'(x,\omega))^2+\omega^2 (u'(x,\omega))^2\right]
\end{equation}
where $u'(x,\omega)$ is the real part of $u(x,\omega)$. The spatial
Fourier transform
\begin{equation}
    u'(x,\omega)=\frac{1}{(2\pi)^{1/2}} \int dq e^{i q x} u'(q,\omega) \;,
\end{equation}
then finally gives
\begin{equation}
    S=\frac{\pi}{2g}\int_{-\infty}^\infty d q \int_{-\infty}^{\infty}
    d\omega \left[ (q^2+\omega^2) | u'(q,\omega)|^2\right] \;.
    \label{constraction}
\end{equation}

Eq. \ref{constraction} allows a straightforward calculation of the
energy of a dislocation near the upper boundary of the slab (see Fig. \ref{fig11}) using
standard tools. We use the correlation function  \cite{Schultz}
\begin{equation}
    G(\tau_m)=\left \langle e^{ i \left[ \phi(x,\tau_m)-\phi(x,-\tau_m)\right] }\right \rangle
\end{equation}
where the boson phase angle $\phi$ is conjugate to $du/dx$. Here $m$
is, as before, the length of the unzipped segment and the angular
brackets denote an average with respect to the action Eq.
\ref{constraction}. In terms of this correlation function the
free-energy of the dislocation is given by
\begin{equation}
    {\cal F}_d(\tau_m)=-\frac{T}{2} \ln G(\tau_m) \;,
\end{equation}
where $T$ is the temperature and we have set, as before, the
Boltzmann constant to be $k_B=1$. Note that we are interested in the
correlation function of the unconstrained field $u(x,\tau)$. The
factor of one-half originates from the fact that the we are
interested in the energy up to $\tau=0$ and not between $-m$ and
$m$. Integrating out the field $\phi$ in the standard way it is easy
to show that this expression can be rewritten as
\begin{equation}
    G(\tau)=\left \langle   \exp\left(-\frac{i \pi}{g} \int_{-\tau_m}^{\tau_m}
    \partial_x u(0,\tau) d\tau\right)\right \rangle
\end{equation}
It is then straightforward to find that for large $m$
\begin{equation}
    {\cal F}_d=\frac{T}{4g} \ln(\tau_m/a_0)
\end{equation}
where $a_0$ is a microscopic cutoff. The result turns out to be
identical to that which would have been obtained without the
constraint on the fields.

\section{Calculation of the dislocation energy in the presence of disorder}
\label{App:dislocation1}

As in the clean sample we consider the correlation function
\begin{equation}
    G(\tau_m)=\left \langle   \exp\left(-\frac{i \pi}{g} \int_{-\tau_m}^{\tau_m}
    \partial_x u(0,\tau) d\tau\right)\right \rangle
\end{equation}
Note that in the clean case the method of images guaranteed that the
flux lines exit normal to the interface. However, with disorder this
will be the case only if disorder in the top image plane is
correlated with the disorder in the bottom plane. However, we expect
that at large $\tau_m$ these correlations will not be important and
we will ignore them.

To obtain the free-energy with disorder and sample-to-sample fluctuations we
consider the replicated correlation function
$\overline{G(\tau_m)^n}$. This allows the extraction of the disorder
averaged free-energy from
\begin{equation}
    \overline{{\cal F}_d(\tau_m)}=T\frac{\partial}{\partial n} \overline{G(\tau_m)^n} |_{n=0}
\end{equation}
and the fluctuations in it from
\begin{equation}
    \overline{\delta {\cal F}^2_d(\tau_m)}=T^2\frac{\partial^2}{\partial n^2}
    \overline{G(\tau_m)^n} |_{n=0}-T^2\left(\frac{\partial}{\partial n} \overline{G(\tau_m)^n} |_{n=0} \right)^2
\end{equation}
Using standard methods it is straightforward to obtain
\begin{eqnarray}
\overline{G(\tau_m)^n} &=&\exp \left( -\frac{1}{2g^2} \left[ n
\int_0^\infty dl g (1-J_0(\tau_m e^{-l}/a_0)) \right.\right. \nonumber \\
&+& \left.\left. n^2 \int_0^\infty dl \kappa(l) (1-J_0(\tau_m
e^{-l}/a_0))   \right]    \right)    \;,
\label{genfunction}
\end{eqnarray}
where $a_0$ is a microscopic cutoff. The Bessel function $J_0(x)$
appearing in Eq. \ref{genfunction} and below is non-universal and
depends on the details of the cutoff procedure. Combining
this with the solutions of the flow equations yields
\begin{equation}
    \overline{{\cal F}_d(\tau_m)}=\frac{T}{4g} \ln(\tau_m/a_0) \;.
\end{equation}
Here we have included a factor of one-half since vortices exist in the lower half plane. The disordered free energy of the dislocation is the
same as that of a dislocation in the clean sample. Similarly one can
obtain for the free-energy fluctuations of the dislocation
\begin{equation}
    \overline{\delta {\cal F}^2_d(\tau_m)}= T^2\frac{\kappa(\infty)}{8g^2} \ln(\tau_m/a_0)
    \label{fluctg1}
\end{equation}
for $g>1$ and
\begin{equation}
    \overline{\delta {\cal F}^2_d(\tau_m)}=T^2\frac{(1-g)^2}{4} \ln^2(\tau_m/a_0)
    \label{fluctg2}
\end{equation}
for $g<1$.

As discussed in Sec.~\ref{3c} a crucial input from the above calculation is that
the free-energy fluctuations grow at most logarithmically in
$\tau_m$.


\begin{thebibliography}{99}
\bibitem{footnote} We note however, that because the relation
(\ref{fb}) is not exact for higher dimensional defects, there is no
symmetry between unzipping and rezipping transitions, i.e.
$\tau_m\to L-\tau_m$ and $f-f_c\to f_c-f$ is the symmetry of the
problem only for 1D columnar defects. See also
Sec.~\ref{sec:numerics}.
\end{thebibliography}
\end{document}